\documentclass{sig-alternate}

\usepackage{graphicx}
\usepackage{url}
\usepackage{amsmath,amssymb,times,mathptmx,verbatim}

\newcommand{\beq}{\begin{equation}}
\newcommand{\eeq}{\end{equation}} 
\def\bearn{\begin{eqnarray*}}
\def\eearn{\end{eqnarray*}}
\def\barr{\begin{array}}
\def\earr{\end{array}}
  
\def\bt{BitTorrent }
\def\btns{BitTorrent}
\def\p2p{peer-to-peer}

\newcommand{\urlsamefont}[1]
{
\urlstyle{same}\url{#1}
}

\begin{document}
\conferenceinfo{SIGMETRICS'07,} {June 12--16, 2007, San Diego, California, USA.}
\CopyrightYear{2007}
\crdata{978-1-59593-639-4/07/0006} 
\toappear{\copyright ACM, 2007. This is the author's version of the work. It
  is posted here by permission of ACM for your personal use. Not for
  redistribution. The definitive version was published in
  Proc. of ACM SIGMETRICS'07, June 12--16, 2007, San Diego, California, USA.} 

\title{Clustering and Sharing Incentives in BitTorrent Systems}

\numberofauthors{2}
\author{
\alignauthor Arnaud Legout\\
\affaddr{I.N.R.I.A.}\\
\affaddr{Sophia Antipolis, France}\\
\email{arnaud.legout@sophia.inria.fr}
\alignauthor Nikitas Liogkas, Eddie Kohler,\\ Lixia Zhang\\
\affaddr{University of California, Los Angeles}\\
\affaddr{Los Angeles, CA USA}\\
\email{\{nikitas, kohler, lixia\}@cs.ucla.edu}
}

\maketitle

\begin{abstract}
Peer-to-peer protocols play an increasingly instrumental role in 
Internet content distribution. It is therefore important to 
gain a complete understanding of how these protocols behave in practice 
and how their operating parameters affect overall system performance. 
This paper presents the first detailed experimental investigation of 
the peer selection strategy in the popular BitTorrent protocol.
By observing more than 40 nodes in instrumented private torrents, we 
validate three protocol properties that, though believed to hold, have
not been previously demonstrated experimentally: the clustering of
similar-bandwidth peers, the effectiveness of BitTorrent's sharing
incentives, and the peers' high uplink utilization.
In addition, we observe that BitTorrent's modified choking algorithm
in seed state provides uniform service to all peers, and that an
underprovisioned initial seed leads to absence of peer clustering
and less effective sharing incentives.
Based on our results, we provide guidelines for seed provisioning 
by content providers, and discuss a tracker protocol extension that 
addresses an identified limitation of the protocol.

\end{abstract}

\category{C.2.2}{Computer-Communication Networks}{Network Protocols}
\category{C.2.4}{Computer-Communication Networks}{Distributed Systems}
\category{C.4}{Performance of Systems}{}
\terms{Algorithms, Measurement, Performance}
\keywords{BitTorrent, choking algorithm, clustering, incentives, seed provisioning}

\section{Introduction}
In just a few years, peer-to-peer content distribution has come 
to generate a significant portion of the total Internet traffic~\cite{karagiannis04}. 
The widespread adoption of such protocols for delivering large data volumes
in a global scale is arguably due to their scalability and robustness properties.
Understanding the mechanisms that affect the performance of such protocols
and overcoming the existing shortcomings will ensure the continued success of 
\p2p data delivery. To that end, this paper presents a detailed experimental 
study of the peer selection strategy in BitTorrent, one of the most popular 
\p2p content distribution protocols.

Recently, researchers have formulated analytical models for the problem of 
efficient data exchange among peers, and measurement studies using actual download 
traces have attempted to shed light into the success of BitTorrent. However, 
certain properties of these studies have interfered with their accurate evaluation
of the dynamics of \bt algorithms and their impact on overall system performance.  
For example, analytical models can provide valuable insight, but they are typically 
based on unrealistic assumptions, such as giving all participants global system 
knowledge~\cite{qiu04}; actual download traces may differ substantially from the 
their predictions~\cite{guo05, pouwelse05}. Furthermore, most measurement studies 
have evaluated peers connected to public \textit{torrents}---BitTorrent download 
sessions~\cite{guo05, izal04, pouwelse05}. They provide detailed data about the 
overall behavior of deployed BitTorrent systems, however, the inherent limitations 
in collecting per-peer information in a public torrent obstructs the understanding 
of individual peer decisions during the download. Legout \textit{et al.}~\cite{legout06} 
recently attempted to evaluate those decisions, but only from the viewpoint of a single peer. 

To overcome these limitations, we conduct extensive experiments on a private testbed 
and collect data from all peers in a controlled environment. 
In particular, we focus on the so-called \textit{choking algorithm} for peer selection, 
which may be the driving factor behind BitTorrent's high performance~\cite{cohen03}.  
This approach allows us to examine the behavior of individual peers under a microscope 
and observe their decisions and interactions during the download. 

Our main contribution is to demonstrate that the choking algorithm facilitates the 
formation of clusters of similar-bandwidth peers, ensures effective sharing incentives 
by rewarding peers who contribute data to the system, and maintains high upload utilization 
for the majority of the download duration. These properties have been hinted at in previous 
work; this study constitutes their first experimental validation.
We also show that, if the seed is underprovisioned, all peers tend to complete their 
download around the same time, independently of how much they upload. Clusters are no longer
formed, and, interestingly, high-capacity peers assist the seed in disseminating data to 
low-capacity ones, resulting in everyone maintaining high upload utilization.
Finally, based on our observations, we provide guidelines for seed provisioning by 
content providers, and discuss a tracker protocol extension that addresses an identified 
limitation of the protocol, namely the low upload utilization at the beginning of a torrent's 
lifetime.

The rest of this paper is organized as follows. Section~\ref{background} provides a 
description of the BitTorrent protocol and an explanation of the choking algorithm, as 
implemented in the official BitTorrent client.  
Section~\ref{methodology} describes our experimental methodology and the rationale
behind the experiments, while Section~\ref{results} presents our results.
Section~\ref{discussion} discusses seed provisioning guidelines and the proposed 
tracker protocol extension. Lastly, Section~\ref{related} sets this study in the context 
of related work, and Section~\ref{conclusion} concludes.

\section{Background}
\label{background}
BitTorrent is a peer-to-peer content distribution protocol that scales well with 
the number of participating peers. A \bt system capitalizes on the upload capacity 
of each peer in order to increase global system capacity as the number of peers 
increases. A major factor behind \btns's success is a built-in incentives mechanism, 
implemented by its \textit{choking algorithm}, which is designed to encourage peers 
to contribute data. The rest of this section introduces the terminology used in the 
paper and describes \btns's operation in detail, with a particular focus on the 
choking algorithm.

\subsection{Terminology}
The terminology used in the \bt community is not standardized. For the sake of
clarity, we define here the terms used throughout this paper.

\begin{itemize}
\item
\textbf{Torrent}. 
A \textit{torrent} is the set of peers cooperating to download the same
  content using the \bt protocol. 

\item
\textbf{Tracker}. 
The \textit{tracker} is the only centralized component of the system.
  It is not involved in the actual distribution of the content, but it 
  keeps track of all peers currently participating in the download, and it
  collects statistics.

\item
\textbf{Pieces and Blocks}. 
Content transferred using \bt is split into \textit{pieces}, with each piece being 
  split into multiple \textit{blocks}. Although blocks are the transmission unit, 
  peers can only share complete pieces with others.

\item
\textbf{Metainfo file}. 
The \textit{metainfo file}, also called a torrent file, contains all the 
  information necessary to download the content and includes the number of pieces, 
  SHA-1 hashes for all the pieces that are used to verify received data, and the IP
  address and port number of the tracker.

\item
\textbf{Interested and Choked}. 
We say that peer $A$ is \textit{interested} in peer $B$ when $B$ has pieces of the
  content that $A$ does not have. Conversely, peer $A$ is \textit{not interested}
  in peer $B$ when $B$ only has a subset of the pieces of $A$. We also say that 
  peer $A$ is \textit{choked} by peer $B$ when $B$ decides not to send any data to 
  $A$. Conversely, peer $A$ is \textit{unchoked} by peer $B$ when $B$ is willing to 
  send data to $A$. 
  Note that this does not necessarily mean that peer $B$ is uploading data to 
  $A$, but rather that $B$ will upload to $A$ if $A$ issues a data request.
 
\item
\textbf{Peer Set}. 
Each peer maintains a list of other peers to which it has open TCP connections.
  We call this list the \textit{peer set}, and it is also known as the neighbor set.

\item 
\textbf{Local and Remote Peers}. 
When describing the choking algorithm, we take the viewpoint of a single peer, which 
  we call the \textit{local peer}. 
  We refer to the peers in the local peer's peer set as \textit{remote peers}.

\item
\textbf{Leecher and Seed}. 
A peer can be in one of two states: the \textit{leecher} state, when it is still
  downloading pieces of the content, and the \textit{seed} state, when it has all 
  the pieces and is sharing them with others. 

\item
\textbf{Initial Seed}. 
The \textit{initial seed} is the first peer that offers the content for download.
  There can be more than one initial seeds. In this paper, however, we only consider
  the case of a single initial seed. 

\item
\textbf{Rarest-First Algorithm}. 
The \textit{rarest-first algorithm} is the piece selection strategy used 
  by \bt clients. It is also known as the \textit{local} rarest-first algorithm 
  since it bases the selection on the available information locally at each peer. 
  Peers independently maintain a list of the pieces each of their remote peers has and 
  build a \textit{rarest-pieces set} containing the indices of the pieces with the least 
  number of copies. This set is updated every time a remote peer announces that it acquired 
  a new piece, and is used by the local peer to select the next piece to download.

\item
\textbf{Choking Algorithm}. 
The \textit{choking algorithm}, also known as the \textit{tit-for-tat algorithm}, is the 
  peer selection strategy used by \bt clients. We provide a detailed description of this
  algorithm in Section~\ref{choking_algorithm}.  

\item \textbf{Official \bt Client}. The official \bt client~\cite{btsite}, also known 
  as the \textit{mainline} client, was the first \bt implementation and was initially
  developed by Bram Cohen, BitTorrent's creator.
\end{itemize}

\subsection{BitTorrent Operation}
Prior to distribution, the content is divided into multiple pieces, and each piece into 
multiple blocks. The metainfo file is then created by the content provider.
To join a torrent, a peer $P$ retrieves the metainfo file out of band, usually from a 
well-known website, and contacts the tracker that responds with a peer set of 
randomly selected peers, possibly including both seeds and leechers. $P$ then starts 
contacting peers in this set and requesting different pieces of the content.

Most clients nowadays use the rarest-first algorithm for piece selection. In this manner,
peer selects the next piece to download from its rarest-pieces set. A
local peer determines which 
pieces its remote peers have based on \textit{bitfield} messages exchanged upon establishing 
new connections, which contain a list of all the pieces a peer has. Peers also send 
\textit{have} messages to everyone in their peer set when they successfully receive and 
verify a new piece. 

A peer uses the choking algorithm to decide which peers to exchange data with. 
The algorithm generally gives preference to those peers who upload data at high rates.  
Once per \textit{rechoke period}, typically set to ten seconds, a peer re-calculates the 
data receiving rates from all peers in its peer set. It then selects the fastest ones, 
a fixed number of them, and uploads only to those for the duration of the period. In 
BitTorrent parlance, a peer unchokes the fastest uploaders via a \textit{regular unchoke}, 
and chokes all the rest.  
In addition, it unchokes a randomly selected peer via a so-called  
\textit{optimistic unchoke}. The logic behind this is explained in detail in 
Section~\ref{choking_algorithm}.

Seeds, who do not need to download any pieces, follow a different unchoke strategy.  
Most implementations dictate that seeds unchoke those leechers that \emph{download} data
at the highest rates, in order to better utilize seed capacity in disseminating the content 
as efficiently as possible.  
However, the official BitTorrent client recently introduced a modified unchoke algorithm in seed 
state, in version 4.0.0. We perform the first detailed experimental evaluation 
of this modified algorithm and show that it enables a more uniform utilization of the 
seed bandwidth across all leechers.

\subsection{Choking Algorithm}
\label{choking_algorithm}
We now describe the choking algorithm in detail as implemented in the official client, 
version 4.0.2. The algorithm was initially introduced to foster a high level of data exchange 
reciprocation and is one of the main factors behind BitTorrent's fairness model: 
peers that contribute data to others at high rates should receive high download throughput, 
and \textit{free-riders}, peers that do not upload, should be penalized by being unable to
achieve high download rates. 
It is worth noting that, although the algorithm has been shown to perform well in a variety 
of scenarios, it has recently been found that it does not completely eliminate 
free-riding~\cite{liogkas07, locher06, sirivianos07}.
In particular, a peer may improve its download rates by downloading from seeds, 
acquiring a large view of the peers in the torrent, or benefiting from many optimistic unchokes.
We discuss this issue further in Section~\ref{sec:provi-sharing-incentives}. 

As we noted earlier, the choking algorithm is different for leechers and seeds. When in leecher 
state, a peer $P$ unchokes a fixed number of remote peers. Unless specified explicitly by the 
user, this number of parallel uploads is determined by $P$'s upload bandwidth. For example, for 
an upload limit greater than or equal to 15 kB/s but less than 42 kB/s this number is set to 4.
For generality, in the following we assume that the number of parallel uploads is set to $n$. 

In leecher state, the choking algorithm is executed periodically at every rechoke
period, i.e., every ten seconds, and in addition, whenever an unchoked and interested 
peer leaves the peer set, or whenever an unchoked peer switches its interest state.  
As a result, the time interval between two executions of the algorithm can sometime be 
shorter than a rechoke period. 
Every time the choking algorithm is executed, we say that a new \textit{round} starts, 
and the following steps are taken.
\begin{enumerate}
\item \label{step:l1} The local peer orders interested remote leechers according to 
  the rates at which it received data from them, and ignores leechers that have not 
  sent any data in the last thirty seconds.  These so-called \textit{snubbed} peers are 
  excluded from consideration in order to guarantee that only contributing peers are unchoked.
\item \label{step:l2} The $n-1$ leechers with the highest rates are unchoked via a 
  \textit{regular unchoke}.
\item \label{step:l3} In addition, every three rounds, an interested
  candidate peer is chosen \emph{at random} to be unchoked via an
  \textit{optimistic unchoke}. If this peer is not unchoked via a
  regular unchoke, it is unchoked via an optimistic unchoke and the
  round completes.  If this peer is already unchoked via a regular unchoke,
  a new candidate peer is chosen \emph{at random}.
  \begin{enumerate}
  \item \label{step:l3a} If the candidate peer is interested in the local peer,
    it is unchoked via an optimistic unchoke and the round completes.
  \item \label{step:l3b} Otherwise, the candidate peer is unchoked anyway, and
  step~\ref{step:l3a} is repeated with a new randomly chosen candidate.
  The round completes when an interested peer is found or when there are no more peers
  to choose, whichever comes first.
  \end{enumerate}
\end{enumerate}

Although more than $n$ peers can be unchoked by the algorithm, only $n$ interested 
peers can be unchoked in the same round. Unchoking non-interested peers improves the 
reaction time in case one of those peers becomes interested during the following 
rechoke period; data transfer can begin right away without waiting for the choking 
algorithm to be executed.
Furthermore, optimistic unchokes serve two major purposes. They function as a resource
discovery mechanism to continually evaluate the upload bandwidth of peers in the peer set
in an effort to discover better partners. They also enable new peers that do not have 
any pieces yet to bootstrap into the torrent by giving them some initial pieces without 
requiring any reciprocation.

In the seed state, older versions of the official client, as well as many current
versions of other clients, perform the same steps as in leecher state, with the 
only difference being that the ordering in step~\ref{step:l1} is based on data
transmission rates from the seed, rather than to it. Consequently, peers with high 
download capacity are favored independently of their contribution to the torrent, a 
fact that could be exploited by free-riders~\cite{liogkas07}.

In version 4.0.0, the official client introduced a modified choking algorithm in seed state.
According to this modified algorithm, a seed performs the same fixed number of $n$ 
parallel uploads as in leecher state, but with different peer selection criteria. 
The algorithm is executed periodically at every rechoke period, i.e., every ten seconds,
and in addition, whenever an unchoked and interested peer leaves the peer set, or 
whenever an unchoked peer switches its interest state.  Every time the choking algorithm 
is executed, a new round starts, and the following steps are taken.

\begin{enumerate}
\item \label{step:s1} The local peer orders the interested and \emph{unchoked} remote
  leechers according to the time it has sent them an unchoke message, most recently 
  unchoked peers first. This is the initial time the local peer had unchoked them; if
  the local peer keeps uploading to them for more than one rechoke periods, it does not 
  send them additional unchoke messages.
  This step only considers leechers to which an unchoke message has been sent recently 
  (less than twenty seconds ago) or leechers that have pending requests for blocks (to 
  ensure that they get the requested data as soon as possible). In case of a tie, leechers 
  are ordered according to their download rates from the seed, fastest ones first, just 
  like the old algorithm did.
  Note that, as leechers do not upload anything to seeds, the notion of snubbed peers does 
  not exist in seed state.
\item \label{step:s2} The number of optimistic unchokes to perform \emph{over the duration
  of the next three rechoke periods}, i.e., thirty seconds, is determined using a heuristic.  
  These optimistic unchokes are uniformly spread over this duration, performing $n_o$ 
  optimistic unchokes per rechoke period. 
  Due to rounding issues, $n_o$ can be different for each of the three rechoke periods.
  For instance, when the number of parallel uploads is 4, the heuristic dictates that only 
  two optimistic unchokes be performed in the entire thirty-second period. Thus, one optimistic 
  unchoke is performed during each of the first two periods and none during the last.
\item \label{step:s3} At each rechoke period, the first $n-n_o$ leechers in the 
  list from step~\ref{step:s1} are unchoked via regular unchokes.
\end{enumerate}

Step~\ref{step:s1} includes the key feature of the modified algorithm in seed state. 
On the one hand, leechers are no longer unchoked based on their observed download rates 
from the seed, but mainly based on the last time an unchoke message was sent to them. 
Thus, after a seed has been sending data to a leecher for six
rechoke periods (when the number of parallel uploads is 4),
it will stop doing so and select another leecher to serve.
In this manner, a seed will provide service to all leechers sooner or later, 
preventing any single leecher from monopolizing it.
On the other hand, according to the official client's version notes, this modified choking 
algorithm in seed state also aims to reduce the amount of duplicate data a seed needs 
to upload before it has pushed out a full copy of the content into the torrent. It 
strives to achieve that by keeping leechers unchoked for six rechoke
periods, in order to prevent high leecher turnover from resulting in
the transmission 
of the same pieces to different leechers.
Interestingly, the most recent version of the official client has reverted back to the 
original choking algorithm in seed state. Although the modified version of the algorithm 
we described here is more robust to modified free-riding implementations, it might be 
less efficient in torrents with compliant peers. Since the company behind the official 
client has been targeting legal content distribution, where client alteration would 
arguably be harder, it may aim to optimize the implementation for this scenario.

Some other implementations have included a \textit{super-seeding} feature with similar 
goals, in particular to assist a service provider with limited upload capacity in seeding 
a large torrent. A seed with this feature masquerades as a normal leecher with no data. 
As other peers connect to it, it will advertise a piece that it has never uploaded before
or that is very rare. After uploading this piece to a given leecher, the seed will 
not advertise any new pieces to that leecher until it sees another peer's 'have' message 
for the piece, indicating that the leecher has indeed shared the piece with others.
This algorithm has anecdotally resulted in much higher seeding efficiencies by reducing 
the amount of duplicate pieces uploaded by the seed, and limiting the amount of data sent 
to peers who do not contribute~\cite{btwikispec}. A single seed running in this mode
is rumored to be able to upload a full copy of the content after only uploading 
105\% of the content data volume. 
Since the official client has not implemented this feature, our experiments do not measure 
its effect on the efficiency of the initial seed. We instead measure the number of 
duplicate pieces uploaded when employing the modified choking algorithm in seed state.

\section{Methodology}
\label{methodology}

\subsection{Experimental Setup}
All our experiments were performed in private torrents on the PlanetLab experimental 
platform~\cite{planetlab}. PlanetLab's convenient tools for collecting measurements 
from geographically dispersed clients greatly facilitated our work. For instance, in 
order to deploy and launch BitTorrent clients on PlanetLab nodes, we utilize the 
\textit{pssh} tools~\cite{pssh}. PlanetLab nodes are typically not behind NATs,
so each peer in our experiments can be uniquely identified by its IP address. 

We chose to experiment on private torrents, as opposed to simulation, in order to 
examine both individual peer decisions and the resulting impact on the torrent. 
Although simulation would have enabled us to run many more experiments, it
would have been a difficult task to accurately model the dynamics of a BitTorrent 
system. Private torrents allow us to observe and record the behavior of all peers in
real scenarios. We can also vary experimental parameters, such as peers' upload rate 
limits, which helps us distinguish which factors are responsible for the observed behavior. 

We performed experiments with the different torrent configurations described in
Section~\ref{configurations}. There are no agreed-upon parameters in the BitTorrent
community, so we set our experiment parameters empirically and based on current best 
practice. During each experiment, leechers download a single file of 113 MB that 
consists of 453 pieces, 256 kB each.

All our experiments were performed with peers that do not change their available upload 
bandwidth during the download, or disconnect before receiving a complete copy of the file. 
There is a single initial seed, and in all experiments, all leechers join the torrent 
at the same time, emulating a flash crowd scenario. Although the behavior of the system 
might be different with other peer arrival patterns, we are interested in examining peer 
decisions under circumstances of high load. The initial seed stays connected to the 
torrent for the duration of the experiment, while leechers disconnect immediately after 
completing their download.

We consider both a well-provisioned and an underprovisioned initial seed.
Seed upload capacity has already been shown to be critical to the performance at 
the beginning of a torrent's lifetime, before the seed has uploaded a complete 
copy of the content~\cite{bharambe06, legout06}. However, the impact of an initial 
seed with limited capacity on system properties is not clear.
Nevertheless, appropriate provisioning of initial seeds is of critical
importance to content providers. We attempt to sketch recommendations on this 
issue in Section~\ref{seed-provisioning} based on our experimental results.

The available bandwidth of PlanetLab nodes is relatively high for typical torrents.
We define upload limits on the leechers and seed to model realistic scenarios,
but \emph{do not define any download limits}, nor do we attempt to match our upload 
limits to inherent limitations of PlanetLab nodes.
Thus, we might end up defining a high upload limit on a node that cannot possibly send 
data that fast, due to network or other problems.
Our results include the effects of local network fluctuations, but we 
believe that the conclusions we draw are not predicated on such effects.
Our experiments utilize 41 PlanetLab nodes, of which 2 are located in Canada and the 
rest are spread across the continental United States. We conduct all runs of an 
experiment consecutively in time on the same set of machines.

We collect our measurements using a modified version of the official BitTorrent 
implementation, instrumented to record interesting events and peer interactions.
Our instrumented client, which is based on version 4.0.2 of the official client
(released in May 2005), is publicly available for download~\cite{btinstrumented}.
We collect a log of each message sent or received along with the content of the 
message, a log of each state change, the rate estimates for remote peers used by the 
choking algorithm, and other relevant information, such as the internal states of the 
choking algorithm. Otherwise specified, we run our experiments with
the default client parameters. 

\subsection{Torrent Configurations}
\label{configurations}
We experimented with several torrent configurations. The parameters we changed from 
configuration to configuration are the upload rate limits for the seed and leechers and 
the upload bandwidth distribution of leechers.
As mentioned before, leecher download bandwidth is never artificially limited, 
although local network characteristics may impose an effective upload or download limit.

We ran experiments with the following configurations.
\begin{itemize}
\item \emph{Two-class}. Leechers are divided into two categories with different
  upload limits. This configuration enables us to observe system behavior in highly 
  bipolar scenarios. Our experiments involve similar numbers of slow peers, with 20~kB/s
  upload limit, and fast peers, with 200~kB/s upload limit.

\item \emph{Three-class}. Leechers are divided into three categories with different 
  upload limits. This configuration helps us identify the qualitative behavioral 
  differences of more distinct classes of peers. Our experiments involve similar 
  numbers of slow peers, with 20~kB/s upload limit; medium peers, with 50~kB/s 
  upload limit; and fast peers, with 200~kB/s upload limit.

\item \emph{Uniform-increase}. Upload limits are defined on leechers according
  to a uniform distribution, with a small 5 kB/s step. The slowest leecher 
  has an upload limit of 20 kB/s, the second slowest a limit of 25 kB/s, 
  and so on.  This configuration provides insight into the behavior of torrents
  with more uniform distribution of peer bandwidth.
\end{itemize}

Our graphs in Section~\ref{results} correspond to experiments run with the
three-class configuration, but the conclusions we draw accord well with the 
results of other experiments. We stress distinctions where appropriate.
We also ran preliminary experiments where the initial seed disconnects
after uploading an entire copy of the content, but leechers remain
connected after they complete their download, serving as seeds for a short
time. Peers in these experiments have somewhat lower completion times
thanks to the extra help from leechers in content dissemination,
but appear otherwise similar. 

\subsection{Experiment Rationale}
\label{rationale}
The goal of our experiments is to understand the dynamics of the choking 
algorithm. To that end, we consider four metrics.

\begin{description}
\item[Clustering:] The choking algorithm aims to encourage high peer reciprocation
  by favoring peers who upload. Therefore, we expect that peers will more 
  frequently unchoke other peers with similar upload capacities, since those are 
  the ones that can reciprocate with high enough rates. The rules for peer selection 
  by Qiu \textit{et al.}~\cite{qiu04} also support this hypothesis.
  Consequently, it is expected that the choking algorithm converges towards good 
  clustering shortly after the beginning of the download by grouping together peers 
  with similar upload capacity. This behavior, however, is not guaranteed and has never 
  been previously verified experimentally. Indeed, let's consider a simple example. 
  Peer $A$ will unchoke peer $B$ if $B$ has been uploading data at a high rate to $A$. 
  In order for $B$ to continue uploading to $A$, $A$ should also start sending data to 
  $B$ at a high enough rate. The only way to initiate such a reciprocal relationship 
  is via an optimistic unchoke. Yet, since optimistic unchokes are performed at random, 
  it is not clear whether and when $A$ and $B$ will get a chance to interact.
  Therefore, in order to preserve clustering, optimistic unchokes should successfully 
  initiate interactions between peers with similar upload capacities. In addition, such 
  interactions should persist despite potential disruptions, such as optimistic unchokes 
  by others or network bandwidth fluctuations.

\item[Sharing incentives:] A major goal of the choking algorithm is to give peers an 
  incentive to share data. The algorithm strives to encourage peers to contribute, since 
  doing so will improve their own download rates. We evaluate the effectiveness of these
  sharing incentives by measuring how peers' upload contributions affect their download
  completion time. We expect that the more a peer contributes, the sooner it will
  complete its download. However, we do not expect to observe strict \textit{data volume 
  fairness}, where all peers contribute the same amount of data; peers who upload at high 
  rates may end up contributing more data than others. They should be rewarded though,
  by completing their download sooner.

\item[Upload utilization:] Upload utilization constitutes a reliable metric of efficiency 
  in \p2p content distribution systems, since the total upload capacity of all peers 
  represents the maximum throughput the system can achieve as a whole. As a result, a 
  \p2p content distribution protocol should aim at maximizing peers' upload utilization.
  We are interested in measuring this utilization in BitTorrent systems, and identifying 
  the factors that can adversely affect it.

 \item[Seed service:] The modified choking algorithm in seed state bases its decisions on the time
   peers have been waiting for seed service, in addition to their download rates from the seed. 
   Thus, we expect to see uniform sharing of the seed upload bandwidth among all peers. It should 
   also be impossible for fast leechers to monopolize the seed.
\end{description}

\section{Experimental Results}
\label{results}
We now report the results of representative experiments that demonstrate our 
main observations. For conciseness, we present only results drawn from the
three-class torrent configuration, but our conclusions are consistent
with our observations from other configurations as well.

\subsection{Well-Provisioned Initial Seed}
We first examine a scenario with a well-provisioned initial seed, i.e., a 
seed that can sustain high upload rates. We expect this to be common for 
commercial torrents, whose service providers typically make sure there is 
adequate bandwidth to initially seed the torrent.
An example might be Red Hat distributing its latest Linux distribution. 
Section~\ref{under_provisioned_seed} shows that peer behavior in the presence 
of an underprovisioned initial seed can differ substantially. 

We consider an experiment with a single seed and 40 leechers: 13 slow peers
(20~kB/s upload limit), 14 medium peers (50~kB/s upload limit), and 13 fast
peers (200~kB/s upload limit). The seed, which is represented as peer 41 in 
the following figures, is limited to upload 200~kB/s, as fast as a fast peer.  
Different peer upload limits are defined in order to model different levels 
of contribution.
The results we report are based on thirteen experiment runs. Although the
official BitTorrent implementation would set the number of parallel uploads 
based on the defined upload limit (4 for the slow, 5 for the medium, and 10 
for the fast peers and the seed), we set this number to 4 for all peers, 
which in fact is what most other clients would do. This ensures homogeneous 
conditions in the torrent and makes it easier to interpret the results. 

\subsubsection{Clustering}
\label{clustering-fast}
\begin{figure}[t]
\centering
\includegraphics[width=0.9\columnwidth]
{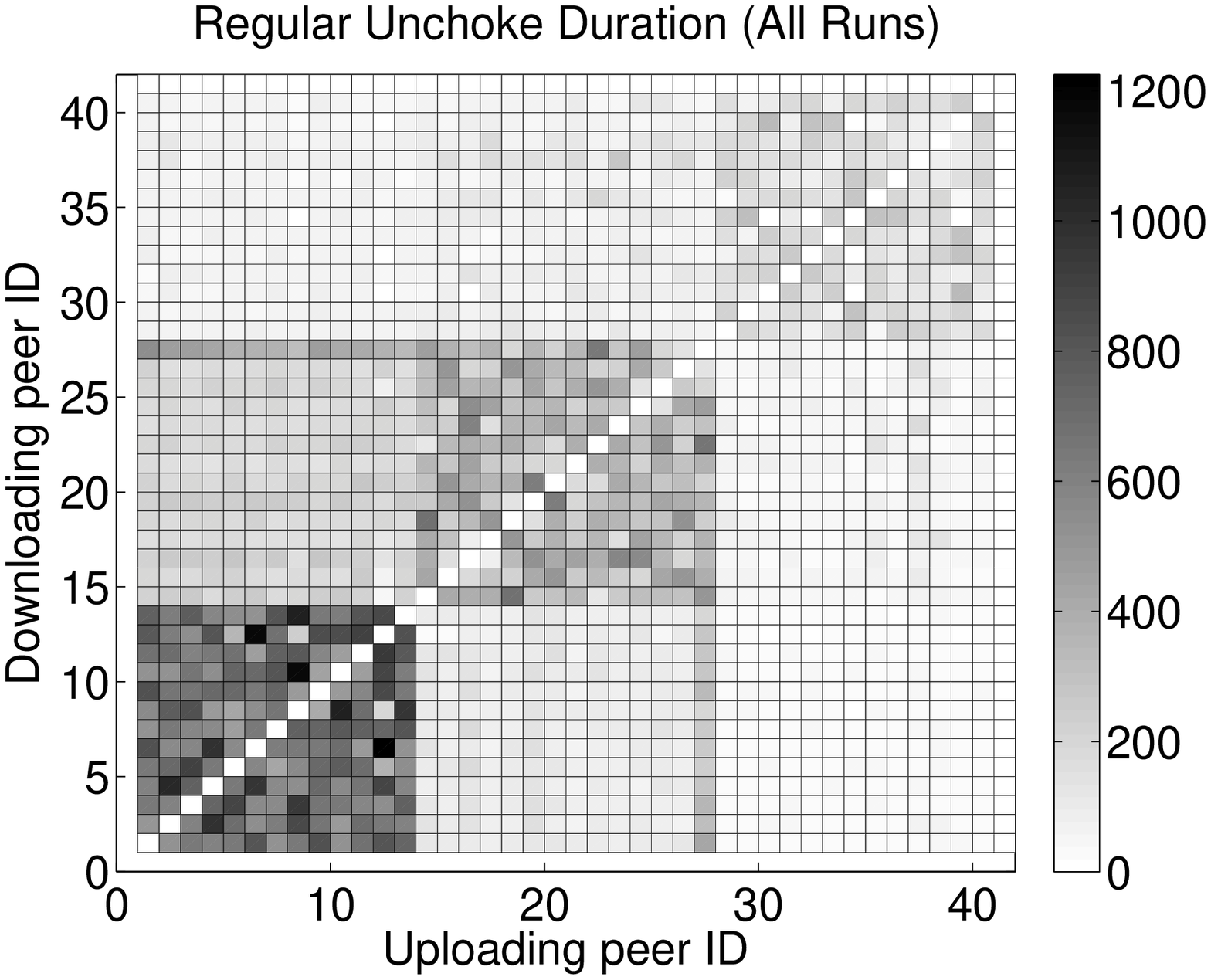}
\caption{\textmd{\textsl{Time duration that peers unchoked each other via a 
  regular unchoke, averaged over all runs. Darker squares represent longer 
  unchoke times (the unit of the color bar on the right is in seconds). 
  Peers 1 to 13 have a 20 kB/s upload limit, peers 14 to 27 have a 50 kB/s 
  upload limit, and peers 28 to 40 have a 200 kB/s upload limit. 
  The seed (peer 41) is limited to 200 kB/s. 
  \emph{The creation of clusters is clearly visible.}}}}
\label{fig:corr-unchoke-upload-onethird-seed-200}
\end{figure}

\begin{figure}[t]
\centering
\includegraphics[width=0.9\columnwidth]
{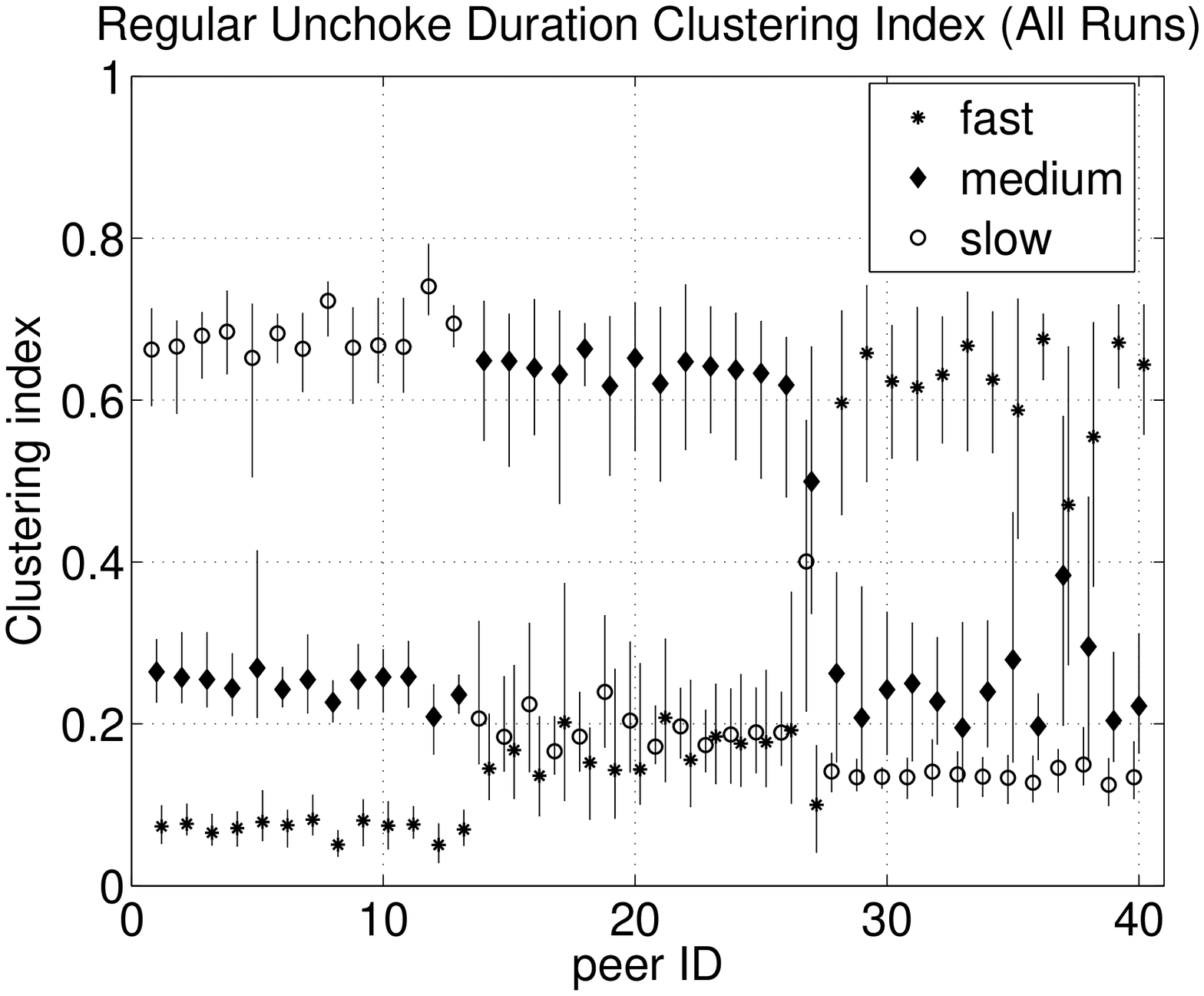}
\caption{\textmd{\textsl{Clustering index for all peers, averaged over all 
  runs, in the presence of a well-provisioned seed.
  Errorbars represent the 10th and 90th percentiles.  
  Peers 1 to 13 have a 20 kB/s upload limit, peers 14 to 27 have a 50 kB/s 
  upload limit, and peers 28 to 40 have a 200 kB/s upload limit. The seed 
  (peer 41) is limited to 200 kB/s. 
  \emph{Peers show a strong preference to unchoke others in the same class.}}}}
\label{fig:clustering-index-onethird-seed-fast}
\end{figure}

\begin{figure}[!ht]
\centering
\includegraphics[width=0.9\columnwidth]
{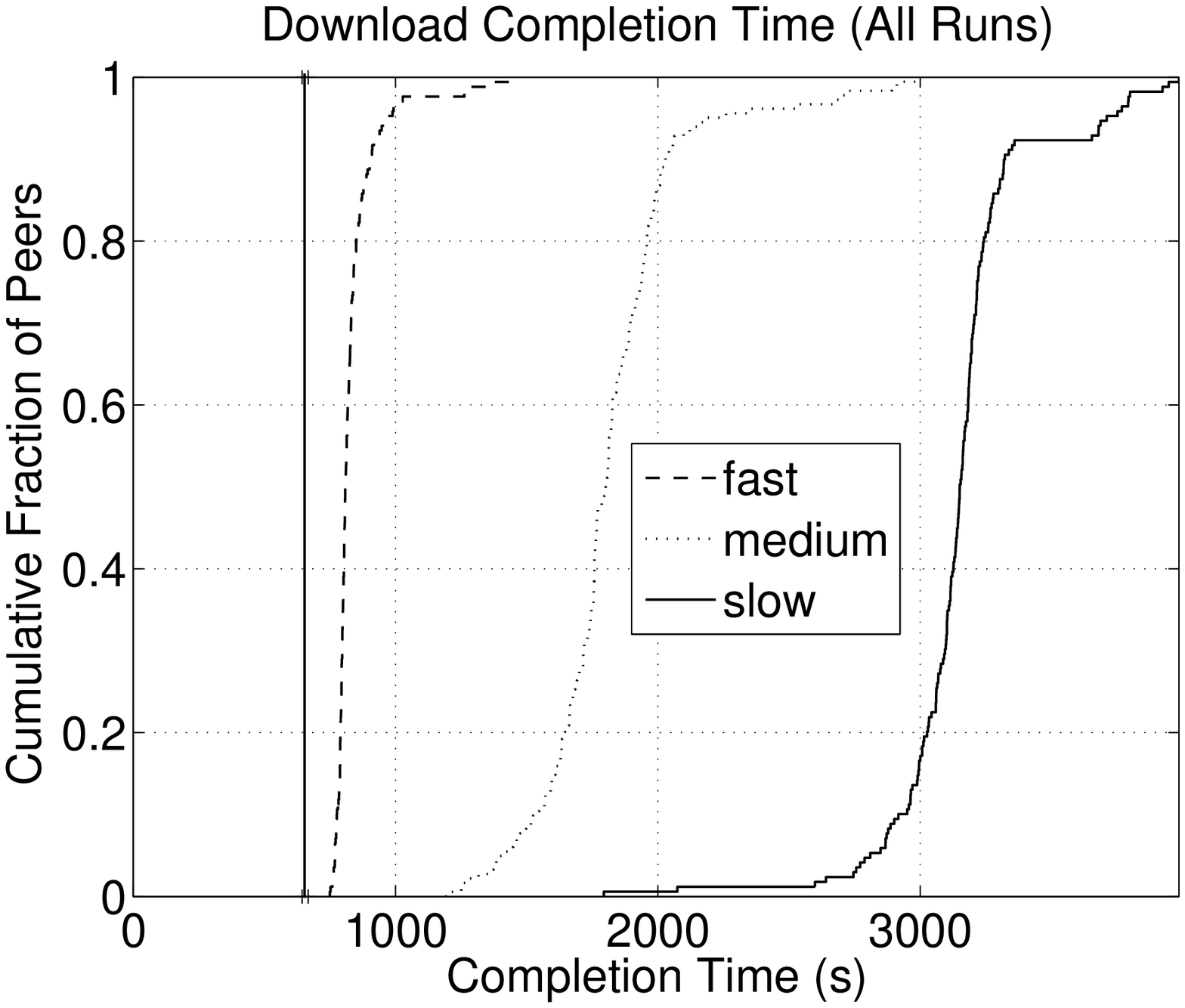}
\caption{\textmd{\textsl{Cumulative distribution of the download completion time 
for the three different classes of leechers, in the presence of a
well-provisioned seed (limited to 200 kB/s), for all runs. 
The vertical line represents the earliest possible time that the download could 
complete. \emph{Fast peers finish much earlier than slow ones.}}}}
\label{fig:completion-cdf-onethird-seed-200-class}
\end{figure}

As explained in Section~\ref{rationale}, we expect to observe clustering
based on peers' upload capacities. 
Figure~\ref{fig:corr-unchoke-upload-onethird-seed-200} demonstrates that 
peers indeed form clusters. The figure plots the total time peers unchoked 
each other via a regular unchoke, averaged over all runs of the experiment. 
It is clear that peers in the same class cluster together, in the sense 
that they prefer to upload to each other. This behavior becomes more apparent 
when considering a metric such as the \emph{clustering index}. We define this 
for a given peer in a given class (fast, medium, or slow) as the ratio of the 
duration of regular unchokes to the peers of its class over the duration of 
regular unchokes to all peers. A high clustering index indicates a strong 
preference to upload to peers in the same class.
Figure~\ref{fig:clustering-index-onethird-seed-fast} plots this index for all
peers and demonstrates that peers prefer to unchoke other peers in their own 
class, thereby forming clusters. Further experiments with upload limits 
following a uniform distribution also show that peers have a clear preference 
for peers with similar upload capacities.

\begin{figure}[t]
\centering
\includegraphics[width=0.9\columnwidth]
{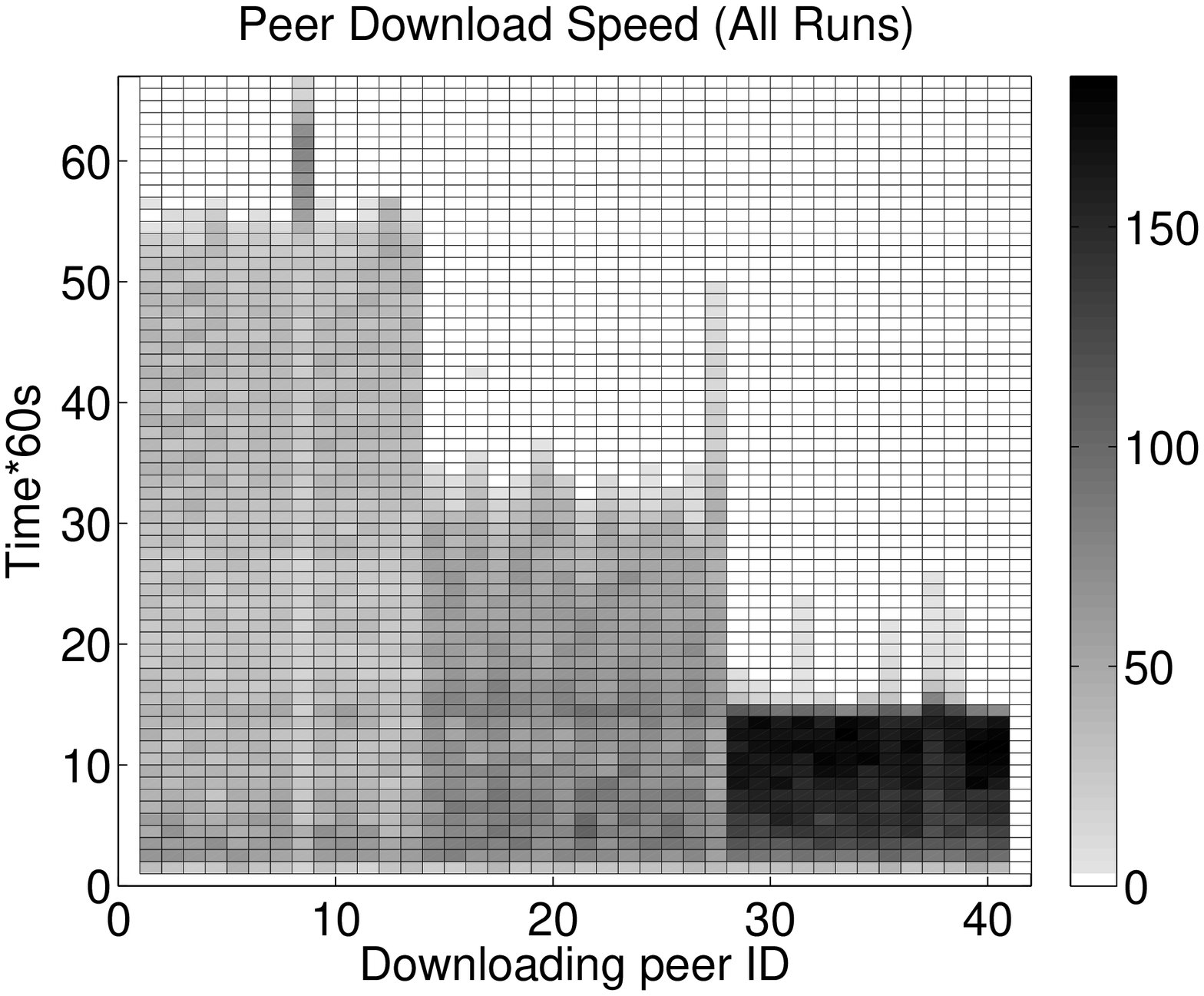}
\caption{\textmd{\textsl{Peer download speeds for all 60-second time intervals 
  during the download, averaged over all runs. Darker rectangles represent 
  higher speeds (the unit of the color bar on the right is in kB/s). 
  Peers 1 to 13 have a 20 kB/s upload limit, peers 14 to 27 have a 50 kB/s 
  upload limit, while peers 28 to 40 have a 200 kB/s upload limit. 
  The seed (peer 41) is limited to 200 kB/s.  
  \emph{Peer 27 achieves lower download rates than other peers in its class,
  while peer 8 is the last one to finish.}}}}
\label{fig:download-speed-onethird-seed-200}
\end{figure}

Although from Figure~\ref{fig:corr-unchoke-upload-onethird-seed-200} it might 
seem that slow peers show a proportionally stronger preference for their own 
class, this is an artifact of the experiment. 
Slow peers take longer to complete their download (as shown in 
Figure~\ref{fig:completion-cdf-onethird-seed-200-class}), and so they perform 
a higher number of regular unchokes on average than fast peers.
Also notice that medium peer 27 interacts frequently with slow peers. 
\emph{This peer's download capacity is inherently limited}, arguably due to 
machine or local network limitations, as seen in 
Figure~\ref{fig:download-speed-onethird-seed-200} that plots observed peer 
download speeds over time. As a result, it stays connected to the torrent even 
after all other peers of its class have completed their download. 
During that last period it has to interact with slow leechers, since those 
are the only ones left. 

Figure~\ref{fig:corr-unchoke-upload-onethird-seed-200} also shows that
reciprocation is not necessarily mutual. Slow peers frequently unchoke
medium peers, but the favor is not returned. Indeed, the slow peers
unchoked medium peers for a total of 501,844 seconds, as shown by the
relatively dark center-left partition. However, the medium peers unchoked slow 
peers for only 273,985 seconds, as shown by the lighter bottom-center. This 
lack of reciprocation is due to the fact that slow peers are of little use to 
medium ones, since they cannot offer high enough upload rates.

In summary, the choking algorithm facilitates clustering, where peers mostly 
interact with others in the same class, with the occasional exception of random
optimistic unchokes.

\begin{figure}[t]
\centering
\includegraphics[width=0.9\columnwidth]
{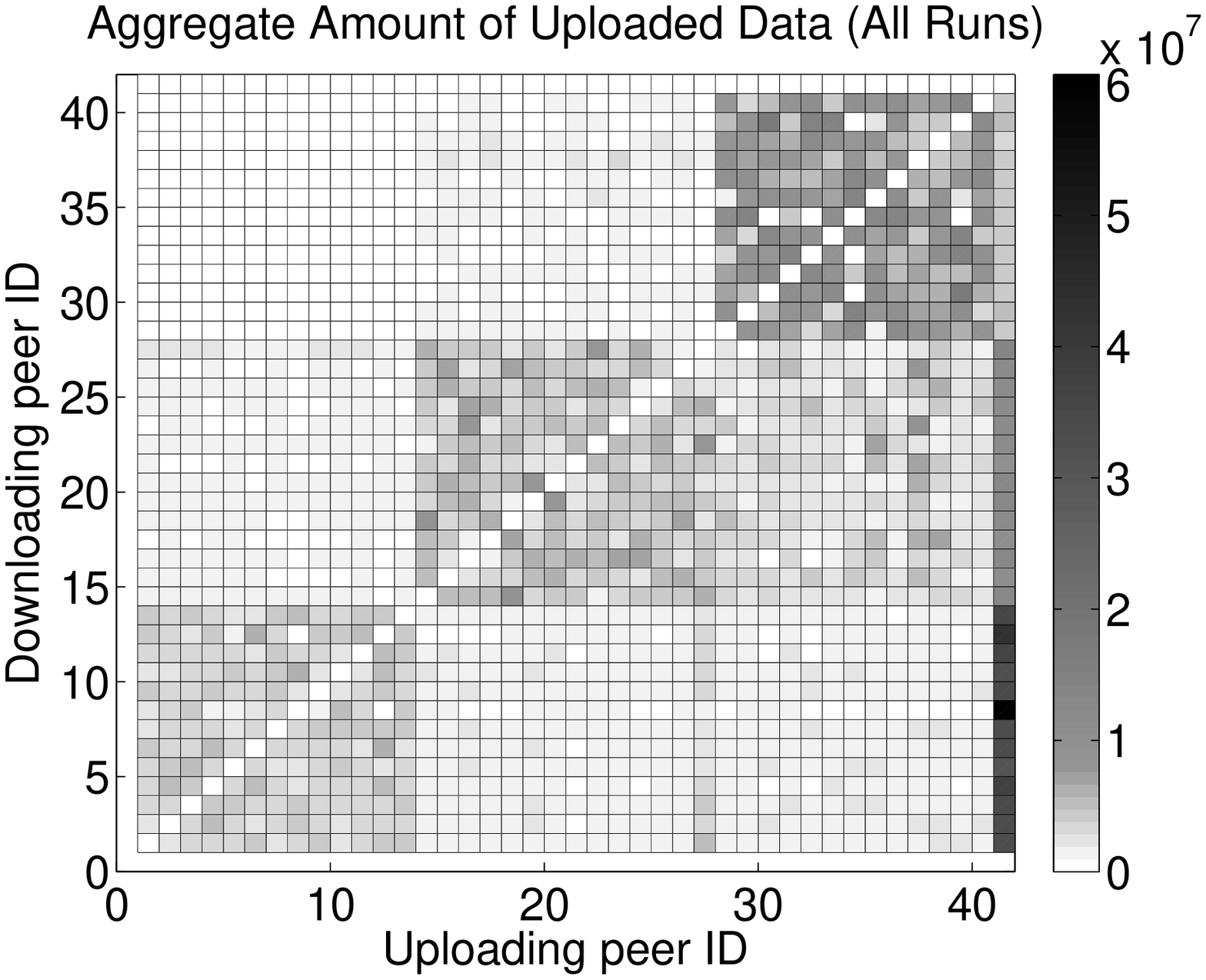}
\caption{\textmd{\textsl{Total number of bytes uploaded by peers to each
  other, averaged over all runs. Darker squares represent more data (the unit 
  of the color bar on the right is in bytes).
  Peers 1 to 13 have a 20 kB/s upload limit, peers 14 to 27 have a 50 kB/s 
  upload limit, and peers 28 to 40 have a 200 kB/s upload limit. 
  The seed (peer 41) is limited to 200 kB/s.  
  \emph{Fast peers upload much more data than the rest.}}}}
\label{fig:agg-amount-bytes-onethird-seed-200}
\end{figure}

\begin{figure}[t]
\centering
\includegraphics[width=0.9\columnwidth]
{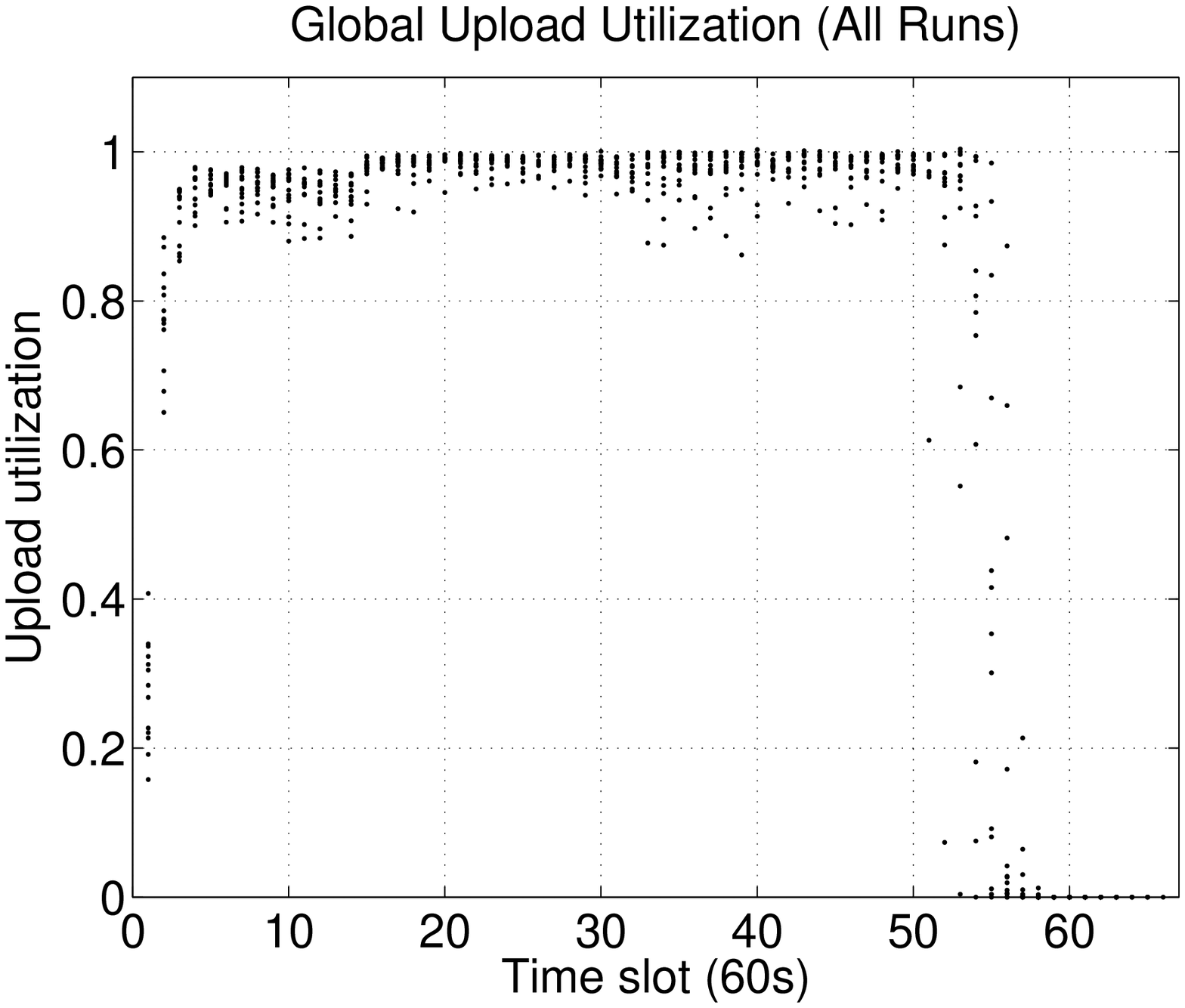}
\caption{\textmd{\textsl{Scatterplot of peers' upload utilization for all 60-second 
  time intervals during the download, in the presence of a well-provisioned seed 
  (limited to 200 kB/s).
  Each point represents the average upload utilization over all peers for a given 
  experiment run.
  \emph{Utilization is kept high during most of the download session.}}}}
\label{fig:up-util-cdf-onethird-seed-200}
\end{figure}

\subsubsection{Sharing Incentives}
\label{sec:provi-sharing-incentives}
We now examine whether BitTorrent's choking algorithm provides effective sharing
incentives, in the sense that a peer who contributes more to the torrent is rewarded 
by completing its download sooner than the rest. 
Figure~\ref{fig:completion-cdf-onethird-seed-200-class} indeed demonstrates 
this to be the case. We plot the cumulative distribution of completion time 
for the three classes of leechers in the previous experiment. The vertical 
line in the figure represents the \emph{optimal completion time}, the earliest 
possible time that any peer could complete its download. This is the time the 
seed finished uploading a complete copy of the content. On average, this time 
is around 650 seconds for the experiment.

Fast leechers complete their download soon after the optimal completion time.  
Medium and, especially, slow leechers take significantly longer to finish.  
Contributing to the torrent enables a leecher to enter the fast cluster and 
receive data at higher rates. This in turn ensures a short download completion 
time. The choking algorithm does indeed foster reciprocation by rewarding
contributing peers. In experiments with upload limits following a uniform 
distribution, the peer completion time is also uniform: completion time 
decreases when a peer's upload contribution increases.  This further indicates 
the algorithm's consistent properties with respect to effective sharing incentives.

Note, however, that this does not imply any notion of data volume fairness.  
Fast peers end up uploading significantly more data than the rest.  
Figure~\ref{fig:agg-amount-bytes-onethird-seed-200}, which plots the 
actual volume of uploaded data averaged over all runs, demonstrates that 
fast peers are the major contributors to the torrent. Most of their bandwidth 
is expended on other fast peers, per the clustering principle. 
Interestingly, the slow leechers end up downloading more data from the seed.
The seed provides equal service to peers of any class, as we show in 
Section~\ref{seed_service-fast}, but slow peers have more
opportunities than others to download from the seed, since  
they take longer to complete.

In summary, BitTorrent  provides effective incentives for peers to contribute, 
as doing so will reward a leecher with significantly higher download rates. 
Recent studies~\cite{liogkas07, locher06, sirivianos07} have shown that limited 
free-riding is possible in BitTorrent under specific circumstances,
although such free-riders do not appear to severely impact the quality of service 
for compliant peers. However, these studies do not significantly challenge the
effectiveness of sharing incentives enforced by the choking algorithm. Although 
free-riding is possible, such peers typically achieve lower download rates 
than they could if they followed the protocol. As a result, if peers wish to 
obtain the highest possible rates, it is in their best interest to conform to 
the protocol.

\subsubsection{Upload Utilization}

We now turn our attention to performance by examining whether the choking 
algorithm can maintain high utilization of peers' upload bandwidth.
Figure~\ref{fig:up-util-cdf-onethird-seed-200} is a scatterplot of such 
utilization in the aforementioned setup. A utilization of 1 represents 
taking full advantage of the available upload capacity. Average utilization for 
each of the thirteen runs is plotted once per minute. The metric is torrent-wide: for 
each minute, we sum the upload bandwidth used by the peers during that minute, 
and divide by the upload capacity available over that minute for all peers still 
connected at the minute's end. The total capacity decreases over time as peers
complete their downloads and disconnect.
Utilization is low at the beginning and the end of the session, but close to 
optimal for the majority of the download. It rises slightly after approximately 
15 minutes, which corresponds to when fast peers leave the torrent. Perhaps the 
four-peer limit on parallel uploads restricts fast peers' utilization. 
In any case, utilization is good overall. 

In summary, the choking algorithm, in cooperation with other BitTorrent mechanisms 
such as rarest-first piece selection, does a good job of ensuring high utilization 
of the upload capacity of leechers during most of the download. 
Low utilization during the startup period may pose a problem for small contents, 
for which it could dominate the total download time. 
We discuss a potential solution to this in Section~\ref{tracker-extension}.

\subsubsection{Seed Service}
\label{seed_service-fast}
\begin{figure}[th]
\centering
\includegraphics[width=0.9\columnwidth]
{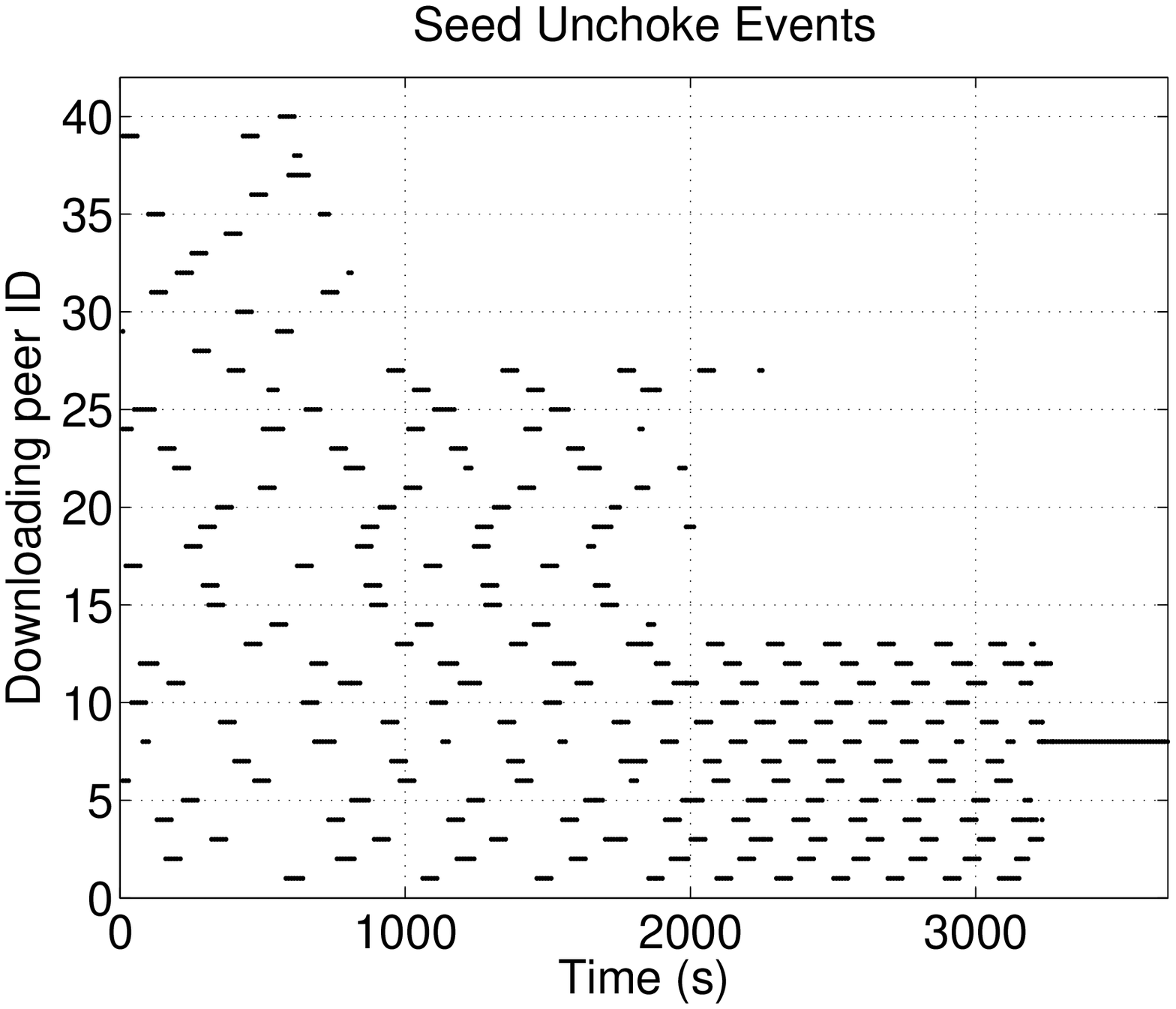}
\caption{\textmd{\textsl{Duration of all unchokes (regular and optimistic) performed 
  by a well-provisioned seed to each peer. Results for a single representative run. 
  Peers 1 to 13 have a 20 kB/s upload limit, peers 14 to 27 have a 50 kB/s upload 
  limit, and peers 28 to 40 have a 200 kB/s upload limit. 
  \emph{The seed (peer 41) provides uniform service to all leechers.}}}}
\label{fig:seed-sku-sru-seed-200}
\end{figure}

\begin{figure}[th]
\centering
\includegraphics[width=0.9\columnwidth]
{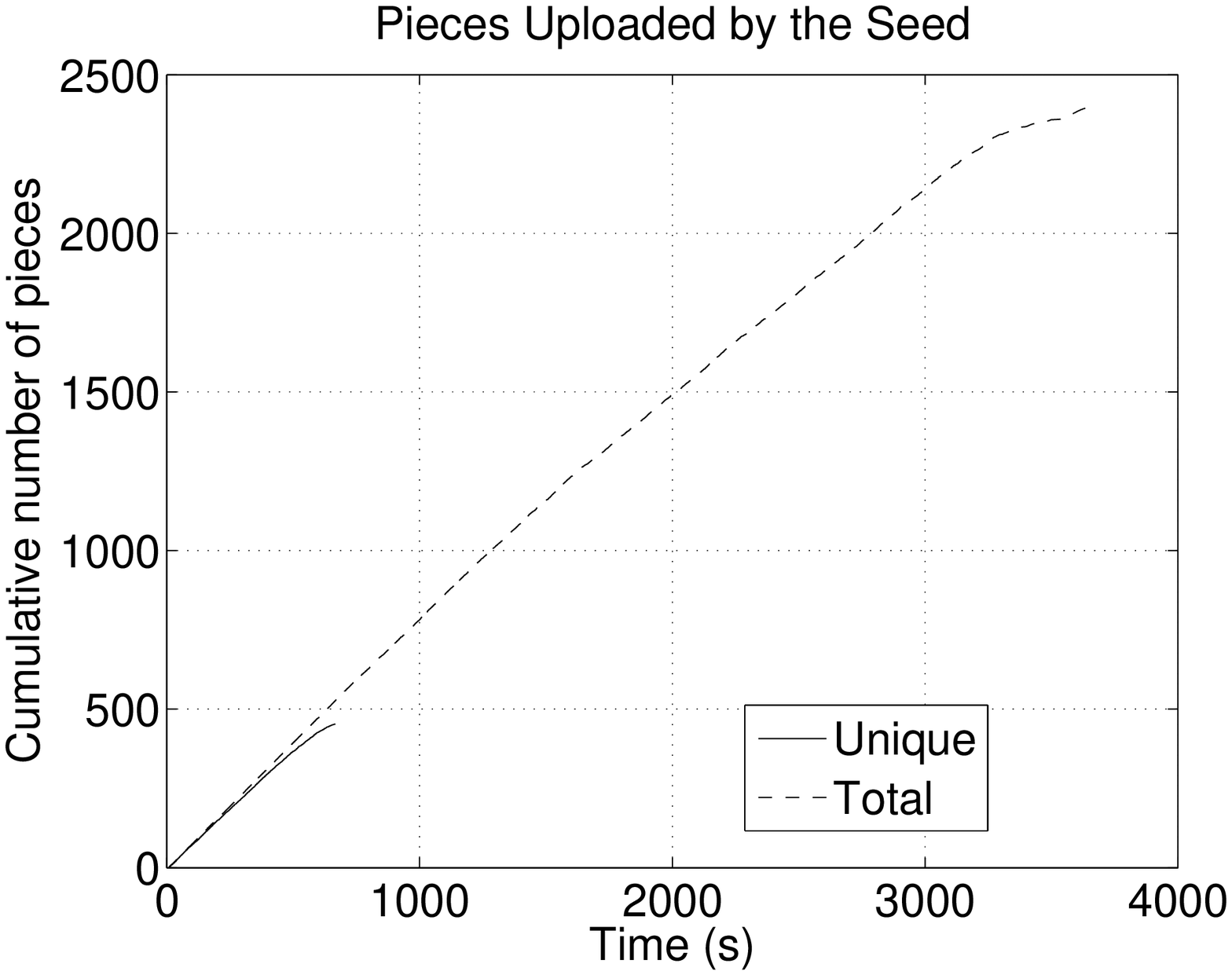}
\caption{\textmd{\textsl{Number of pieces uploaded by the seed (limited to 200 kB/s), 
  for a single representative run. The Unique line represents the pieces that had not 
  been previously uploaded, while the Total line represents the total number of pieces 
  uploaded so far. \emph{We observe a 14\% duplicate piece overhead.}}}}
\label{fig:seed-unique-piece-seed-200}
\end{figure}

The official client introduced a modified choking algorithm in seed state, as described 
in Section~\ref{choking_algorithm}, although it reverted back to the original in the 
most recent version.
The client's version notes claim that the modified algorithm aims to reduce the 
amount of duplicate data a seed needs to upload before it has pushed out a full copy 
of the content into the torrent. We study this modified algorithm for the first time and 
examine this claim. 

Figure~\ref{fig:seed-sku-sru-seed-200} shows the duration of unchokes, both regular and
optimistic, performed by the seed in a representative run of the aforementioned setup. 
Leechers are unchoked in a uniform manner, regardless of upload speed. Fast peers, those 
with higher peer IDs, complete their download sooner, after which time the seed divides 
its upload bandwidth among the remaining leechers. Leecher 8 is the last to complete 
(as shown in Figure~\ref{fig:download-speed-onethird-seed-200}), and receives exclusive 
service from the seed during the end of its download. 
We therefore see that the modified choking algorithm in seed state provides uniform service; 
this is because it bases its unchoking decisions on the time peers have been waiting for 
seed service.
As a result, the risk of fast leechers downloading the entire content and quickly 
disconnecting from the torrent is significantly reduced. Furthermore, this behavior would 
mitigate the effectiveness of exploits that attempt to monopolize seeds~\cite{liogkas07}.

According to anecdotal evidence~\cite{btwikispec}, initial seeds using the old algorithm 
might have to upload 150\% to 200\% of the total content size before other peers become 
seeds. Our experiments show that the modified algorithm avoids this problem.
Figure~\ref{fig:seed-unique-piece-seed-200} plots the number of pieces uploaded by the 
seed during the download session for a representative run. 527 pieces are sent out before 
an entire copy of the content (453 pieces) has been uploaded. Thus, the duplicate piece 
overhead is around 14\%, indicating that the modified choking algorithm in seed state avoids
unnecessarily uploading duplicate pieces to a certain extent. 
This number was consistent across all our experiments, ranging from 11 to 15\%. 
However, to the best of our knowledge, there has been no experimental evaluation 
of the corresponding overhead in the old algorithm, so it is not clear how much of an 
improvement this is.

In any case, 14\% duplication represents an opportunity for improvement.
The official client always issues requests for pieces in the rarest-pieces set in the 
same order. As a result, leechers might end up requesting the same piece from the seed 
at approximately the same time. It would be preferable for leechers to request rarest 
pieces in random order instead.

\subsection{Underprovisioned Initial Seed}
\label{under_provisioned_seed}
We now turn our attention to a scenario with an underprovisioned initial seed and 
demonstrate that the seed upload capacity is critical to performance during the 
beginning of a torrent's lifetime.
The experiment we present here involves a single seed and 39 leechers, 12 slow, 14 medium, 
and 13 fast.  These nodes are different than the nodes used in the previous experiment.
The initial seed, represented as peer 27 in the following figures, is in this case limited 
to 100~kB/s, instead of 200~kB/s. We set the number of parallel uploads again 
to four for the seed and all the leechers. The results we present are based on eight 
experiment runs and are consistent with our observations from experiments with other 
torrent configurations. Peer behavior in the presence of an underprovisioned initial 
seed is substantially different than with a well-provisioned one.

\subsubsection{Clustering}
\begin{figure}[t]
\centering
\includegraphics[width=0.9\columnwidth]
{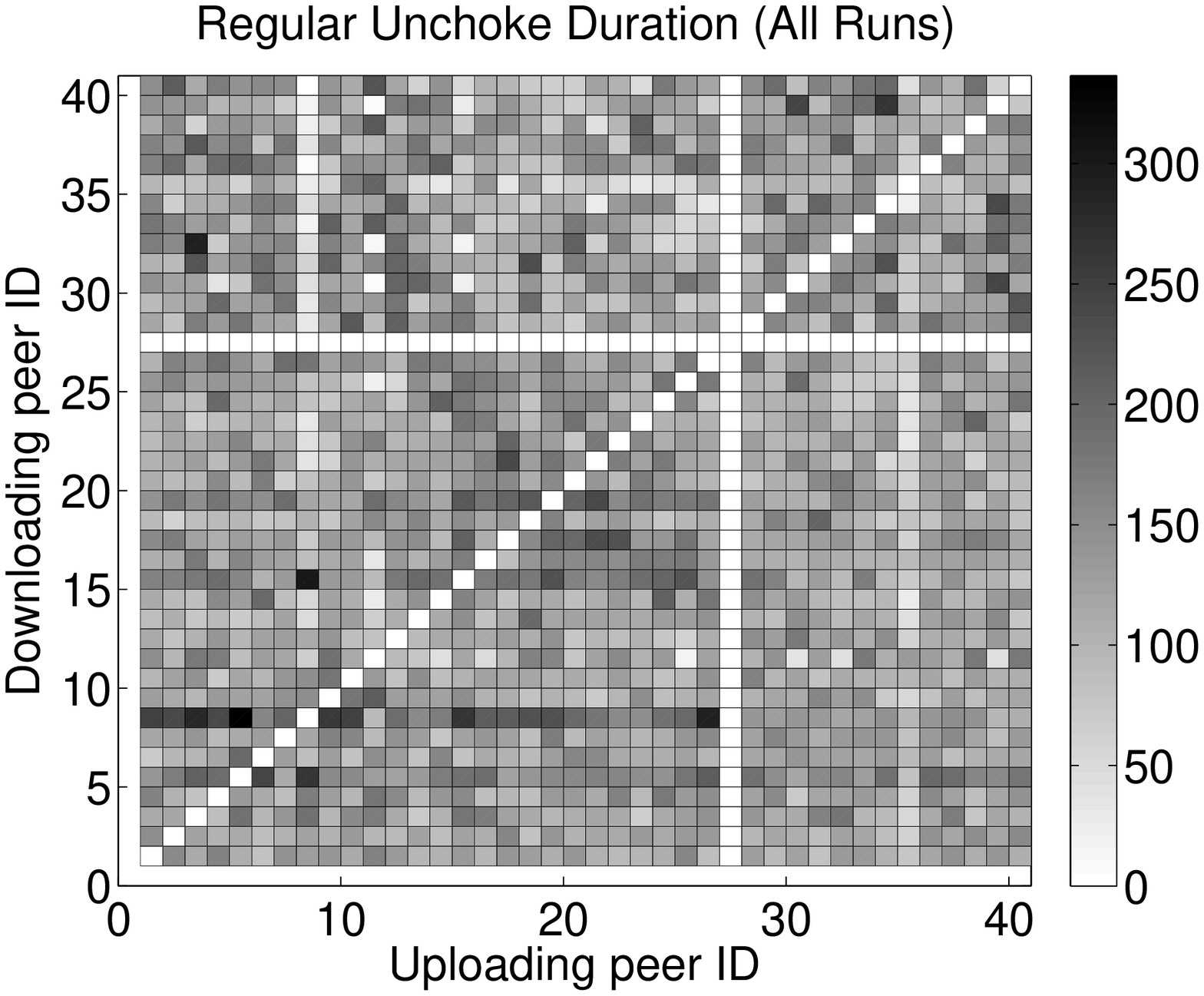}
\caption{\textmd{\textsl{Time duration that peers unchoked each other via a regular unchoke, 
  averaged over all runs. Darker squares represent longer unchoke times (the unit of the 
  color bar on the right is in seconds).
  Peers 1 to 12 have a 20 kB/s upload limit, peers 13 to 26 have a 50 kB/s upload limit, 
  and peers 28 to 40 have a 200 kB/s upload limit. 
  The seed (peer 27) is limited to 100 kB/s.
  \emph{There is no discernible clustering.}}}}
\label{fig:corr-unchoke-upload-onethird-seed-medium}
\end{figure}

\begin{figure}[t]
\centering
\includegraphics[width=0.9\columnwidth]
{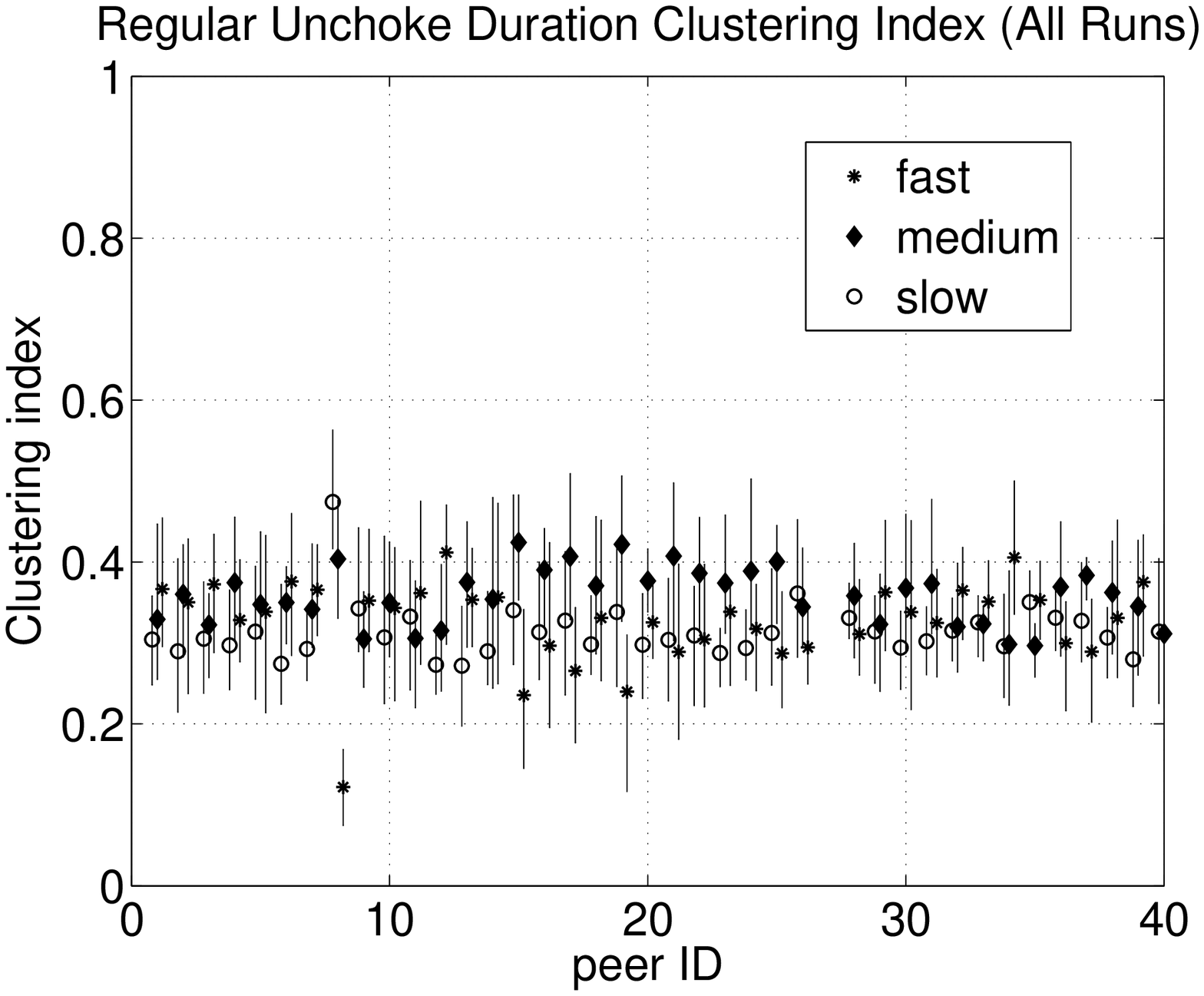}
\caption{\textmd{\textsl{Clustering index for all peers in the presence of an 
  underprovisioned seed, averaged over all runs.
  Errorbars represent the 10th and 90th percentiles.  
  Peers 1 to 12 have a 20 kB/s upload limit, peers 13 to 26 have a 50 kB/s upload limit,
  and peers 28 to 40 have a 200 kB/s upload limit. 
  The seed (peer 27) is limited to 100 kB/s.  
  \emph{Peers do not show a clear preference to unchoke other peers in any particular class.}}}}
\label{fig:clustering-index-onethird-seed-medium}
\end{figure}

\begin{figure}[!ht]
\centering
\includegraphics[width=0.9\columnwidth]
{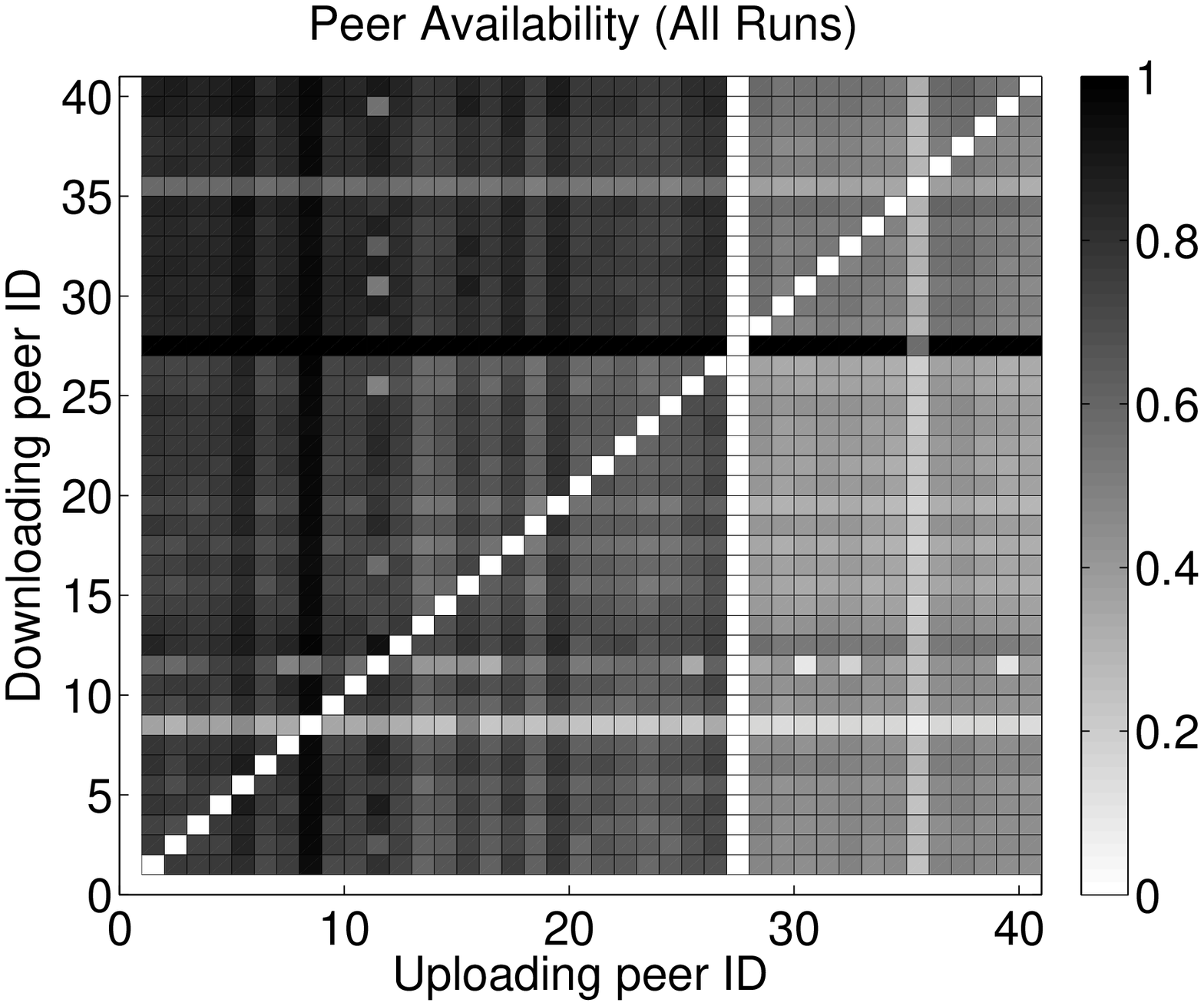}
\caption{\textmd{\textsl{Normalized interested time duration for each peer pair, averaged 
  over all runs. Darker squares represent higher peer availability.  Peers 1 to 12 have a 
  20 kB/s upload limit, peers 13 to 26 have a 50 kB/s upload limit, and peers 28 to 40 have 
  a 200 kB/s upload limit. The seed (peer 27) is limited to 100 kB/s.  
  \emph{Fast peers have poor peer availability to all other peers.}}}}
\label{fig:cumul-interest-run1-seed-medium}
\end{figure}

Figure~\ref{fig:corr-unchoke-upload-onethird-seed-medium} shows the total time peers 
unchoked each other via a regular unchoke, averaged over all runs of the experiment. 
In contrast to Figure~\ref{fig:corr-unchoke-upload-onethird-seed-200}, 
there is no discernible clustering among peers in the same class. The lack of clustering 
in the presence of an underprovisioned initial seed becomes more apparent when considering 
the clustering index metric defined in Section~\ref{clustering-fast}.
Figure~\ref{fig:clustering-index-onethird-seed-medium} shows this metric for all peers. 
They are all similar, indicating a lack of preference to unchoke peers in any particular class. 

Figure~\ref{fig:cumul-interest-run1-seed-medium} attempts to explain this behavior by 
plotting the peer availability of each peer to every other peer, averaged over all 
runs of the experiment. We define the \emph{peer availability} of a downloading peer 
$Y$ to an uploading peer $X$ as the ratio of the time $X$ was interested in $Y$ to 
the time that $Y$ spent in the peer set of $X$. A peer availability of 1 means that
the uploading peer was always interested in the downloading peer, while a peer 
availability of 0 means that the uploading peer was never interested in the downloading peer.

We can see that the fast peers have poor peer availability to all other peers. This is
because the seed is uploading new pieces at a low rate, so even if it uploaded only to 
fast peers, those would quickly replicate every piece as it was completed, remaining 
non-interested for the rest of the time. The same is not true for slow peers, however, 
since they upload even more slowly than the seed. In addition, when a fast leecher is 
unchoked by a slow leecher, it will always reciprocate with high rates, and thereby be 
preferred by the slow leecher. As a result, fast peers will get new pieces even from medium 
and slow peers. In this manner, fast peers prevent clustering by taking up slower peers' 
unchoke slots and thus breaking any clusters that might be starting to form. This prevents 
medium and slow peers from clustering together, even though the seed is fast enough with 
respect to them. Further experiments with other torrent configurations, including one with 
the initial seed further limited to 20~kB/s, confirm this conclusion.

In summary, when the initial seed is underprovisioned, the choking algorithm
does not enable peer clustering. We study in the next section how
this lack of clustering affects the effectiveness of sharing
incentives.

\subsubsection{Sharing Incentives}
\begin{figure}[t]
\centering
\includegraphics[width=0.9\columnwidth]
{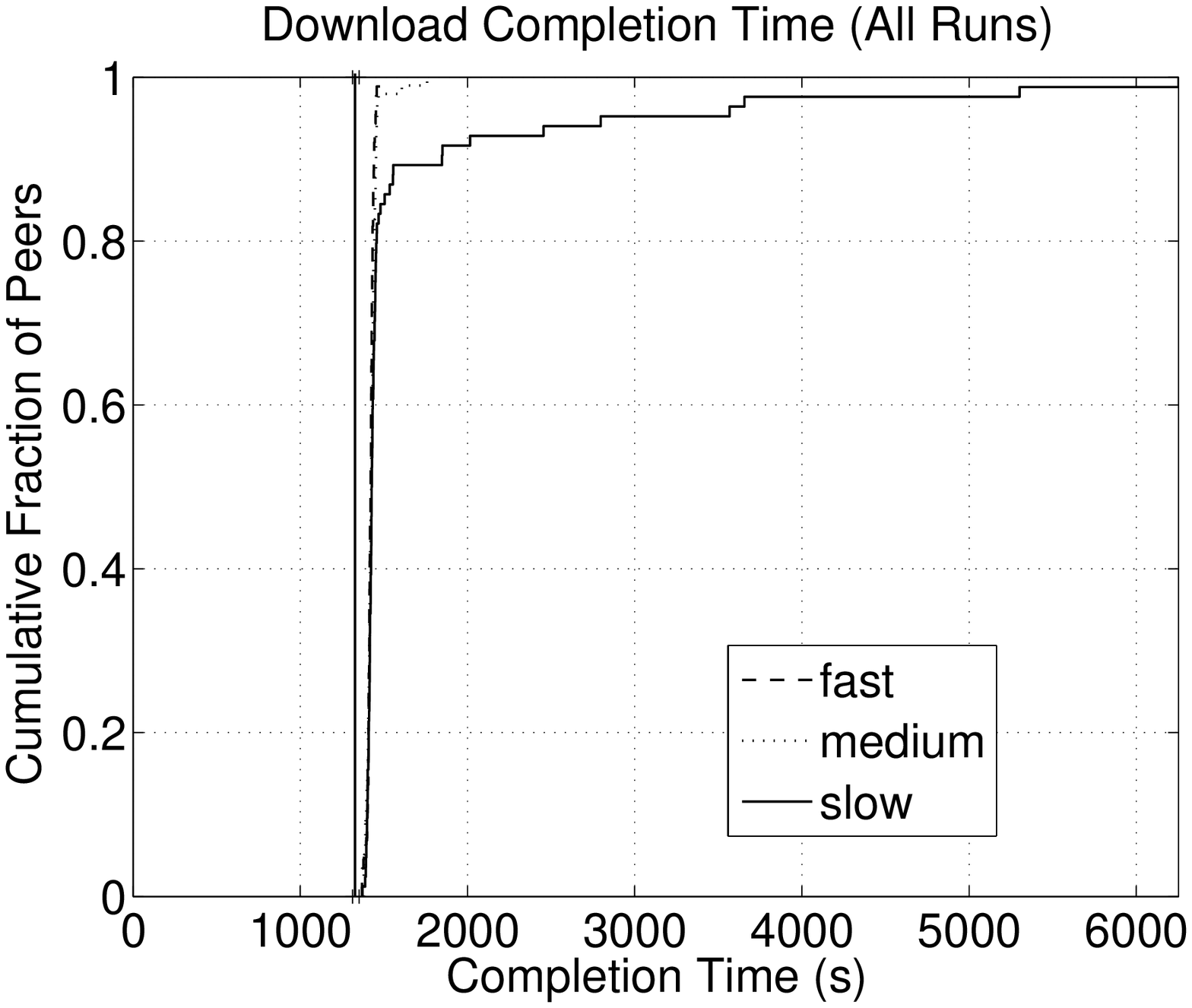}
\caption{\textmd{\textsl{Cumulative distribution of the download completion time for 
  the three different classes of leechers, in the presence of an underprovisioned seed 
  (limited to 100 kB/s), for all runs. 
  The vertical line represents the earliest possible time that the download could complete.  
  \emph{Most peers complete at approximately the same time, regardless of their contribution,
  soon after the seed finishes uploading a complete copy of the content.}}}}
\label{fig:completion-cdf-onethird-seed-medium-class}
\end{figure}

\begin{figure}[t]
\centering
\includegraphics[width=0.9\columnwidth]
{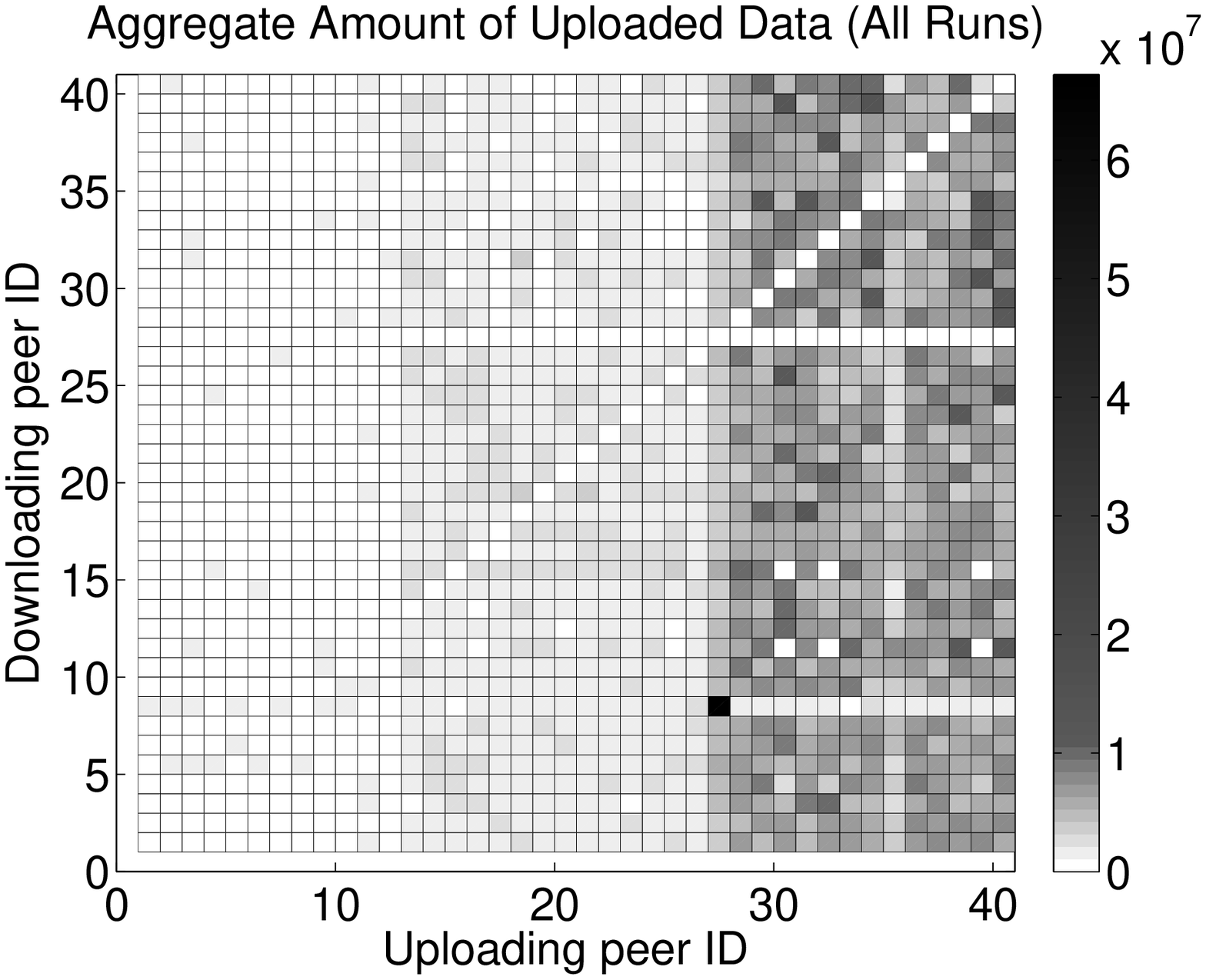}
\caption{\textmd{\textsl{Total number of bytes peers uploaded to each other, averaged over 
  all runs. Darker squares represent more data (the unit of the color bar on the right is in bytes).
  Peers 1 to 12 have a 20 kB/s upload limit, peers 13 to 26 have a 50 kB/s upload limit, 
  and peers 28 to 40 have a 200 kB/s upload limit. The seed (peer 27) is limited to 100 kB/s.
  \emph{Fast peers upload the most data, spreading their bandwidth evenly.}}}}
\label{fig:agg-amount-bytes-onethird-seed-medium}
\end{figure}

\begin{figure}[!ht]
\centering
\includegraphics[width=0.9\columnwidth]
{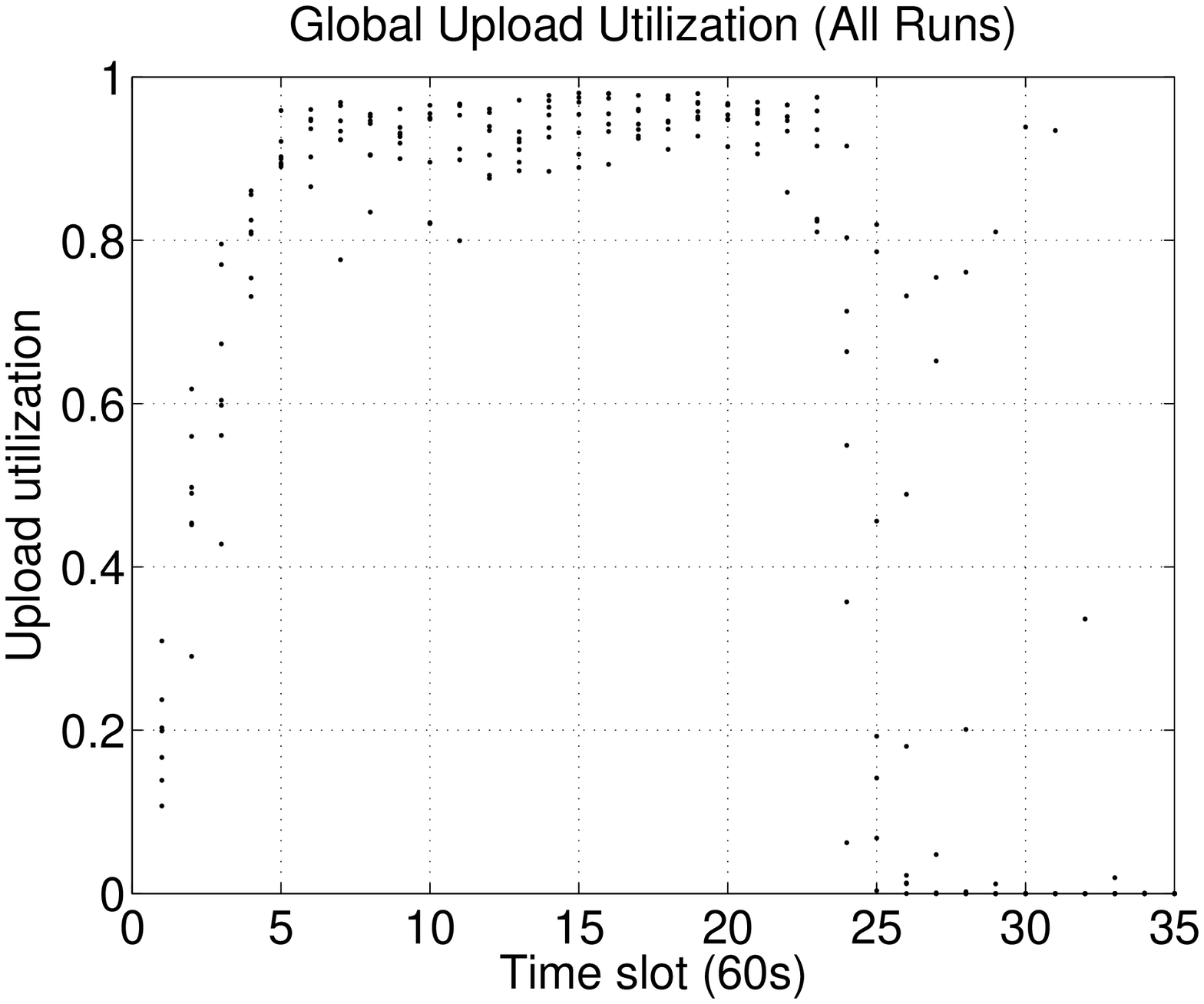}
\caption{\textmd{\textsl{Scatterplot of peers' upload utilization for all 60-second time 
  intervals during the download, in the presence of an underprovisioned seed (limited to 100 kB/s). 
  Each point represents the average upload utilization over all peers for a given experiment run. 
  \emph{Utilization is kept at acceptable levels despite the seed limitation.}}}}
\label{fig:up-util-cdf-onethird-seed-medium}
\end{figure}
We now examine how the lack of clustering affects the effectiveness of sharing incentives. 
In particular, we investigate whether fast peers still complete their download sooner than 
the rest. Figure~\ref{fig:completion-cdf-onethird-seed-medium-class} shows that this is no 
longer the case. Most peers complete their download at approximately the same time. The 
points in the tail of the figure are due to a single slow peer, peer 8, which completed its 
download last in every run. This PlanetLab node has a poor effective download speed 
independently of the choking algorithm, likely due to machine or local network limitations.
All other peers, for all runs, complete their download less than 2,000 seconds after
the beginning of a run. Clearly, seed upload capacity is the performance bottleneck. Once 
the seed finishes uploading a complete copy of the content, all peers complete soon thereafter.
Since uploading data to others does not shorten a peer's completion time, BitTorrent's sharing 
incentives do not seem to be effective in this situation.

Fast peers are again the major contributors in the torrent, but in this case their upload 
bandwidth is expended equally across other fast and slower peers alike.  
Figure~\ref{fig:agg-amount-bytes-onethird-seed-medium}, which plots the amount of uploaded 
data between each peer pair, shows that fast peers made the most contributions, distributing 
their bandwidth evenly to all other peers.

In summary, when the initial seed is underprovisioned, the choking algorithm does 
not provide effective incentives to contribute. Nevertheless, the available upload capacity 
of fast peers is effectively utilized to replicate the pieces being uploaded by the seed.

\begin{figure}[t]
\centering
\includegraphics[width=0.9\columnwidth]
{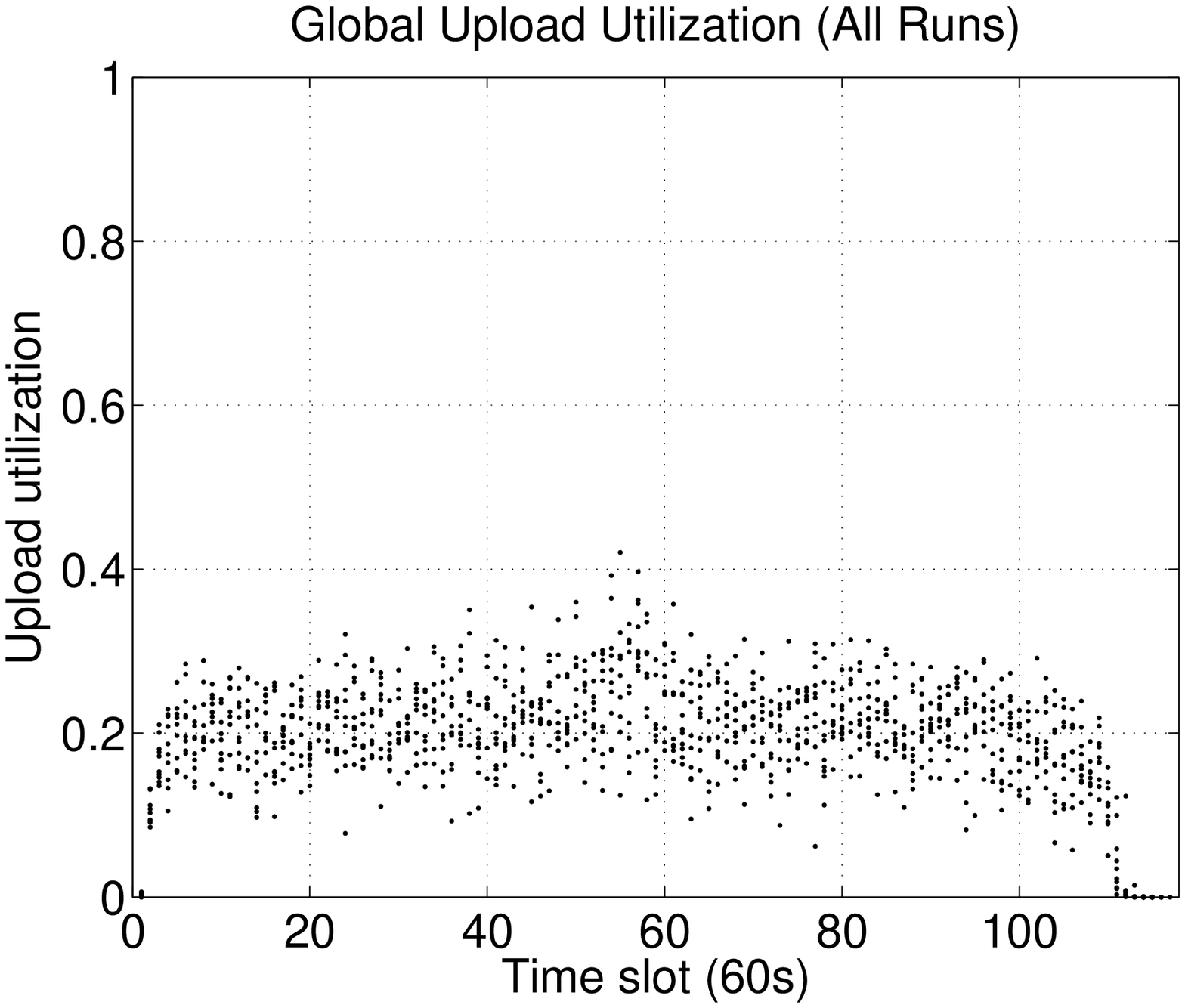}
\caption{\textmd{\textsl{Scatterplot of peers' upload utilization for all 60-second time 
  intervals during the download, in the presence of a severely underprovisioned seed (limited 
  to 20 kB/s). Each point represents the average upload utilization over all peers for a 
  given experiment run. \emph{Utilization is poor when the seed is very slow.}}}}
\label{fig:up-util-cdf-onethird-seed-slow}
\end{figure}
\subsubsection{Upload Utilization}
Interestingly, even with a slow seed, upload utilization remains relatively high, as shown in
Figure~\ref{fig:up-util-cdf-onethird-seed-medium}.
Leechers manage to exchange data productively among themselves once new pieces are downloaded 
from the seed, so that the lack of clustering does not degrade overall performance significantly. 
The BitTorrent design seems to lead the system to do the right thing: fast peers contribute 
their bandwidth to reduce the burden on the initial seed, helping disseminate the available
pieces to slower peers. Although this destroys clustering, it improves overall efficiency, 
which is a reasonable trade-off given the situation.

We also experimented with a seed limited to an upload capacity of 20~kB/s. 
Figure~\ref{fig:up-util-cdf-onethird-seed-slow} shows that, with this extremely low seed 
capacity, there are few new pieces available to exchange at any point in time, and 
each new piece gets disseminated rapidly after it is retrieved from the seed.
The overall upload utilization is now low. Slow peers exhibit slightly higher 
utilization than the rest, since they do not need many available pieces to use up their
available upload capacity. 

In summary, even in situations where the initial seed is underprovisioned, the global
upload utilization can be high. However, our experiments only involve compliant clients, 
who do not try to adapt their upload contributions according to a utility function of the
observed download speed. On the other hand, in an environment with free-riders and an
underprovisioned seed, one might expect  a lower upload utilization due to the lack of 
altruistic peer contributions.

\section{Discussion}
\label{discussion}
We now discuss two limitations of the choking algorithm that we identified through
our experiments: the initial seed upload capacity is fundamental to the proper 
operation of the incentives mechanism, and peers take some time to reach full upload 
utilization at the beginning of the download session.

\subsection{Seed Provisioning}
\label{seed-provisioning}
When the initial seed is underprovisioned, the choking algorithm does not lead to 
the clustering of similar-bandwidth peers. Even without clustering, however, we observed 
high upload utilization.
Interestingly, in the presence of a slow initial seed, the protocol mechanisms lead
the fast leechers to contribute to the download of all other peers, fast or slow, 
thereby improving performance.

However, whenever feasible, one should engineer adequate initial seed capacity 
in order to allow fast leechers to achieve optimal performance. Our results show that 
the lack of clustering occurs when fast peers cannot maintain their interest in other
fast peers. In order to avoid this situation, the initial seed should \emph{at least 
be able to upload data at a speed that matches that of the fastest peers in the torrent}.
This suggestion is of course a rule-of-thumb guideline, and assumes that the service
provider knows a priori the maximum upload capacity of the peers that may join the 
torrent in the future.
In practice, reasonable bounds could be derived from measurements or from an
analysis of deployed network technologies.
Further research is needed to evaluate the exact impact of initial seed capacity.
We are currently developing an analytical model that attempts to express the effect 
of this parameter on peer performance.

\subsection{Tracker Protocol Extension}
\label{tracker-extension}
When a new leecher first joins the torrent, it connects to a random subset 
of already-connected peers that are returned by the tracker. However, in order to
reach its optimal bandwidth utilization, this new leecher needs to exchange data 
with those peers that have a similar upload capacity to itself. If there are few 
such peers in the torrent, it may take some time to discover them, since this has 
to be done via random optimistic unchokes that occur only once every 30 seconds.

Consequently, it might be preferable to utilize the tracker in matching similar-bandwidth 
leechers. In this manner, the duration of the discovery period could decrease and the 
upload utilization would be high even at the beginning of a peer's download.
The new leecher could \emph{report its available upload capacity to the tracker when 
joining the torrent}. This parameter can be configured in the client software, or may 
possibly be the actual maximum upload rate measured during previous downloads. 
The tracker would then reply with a random subset of peers as usual, along with their 
upload capacities. The new leecher could optionally perform optimistic unchokes first 
to peers with similar upload capacity, in an effort to discover the best partners sooner.

Using this new tracker protocol extension, if the peer set contains only a few
leechers with similar upload capacity, they will discover each other quickly.
Leechers should employ some means of detecting and punishing others who lie about their 
available upload capacity. For instance, if a leecher does not respond to an optimistic 
unchoke with an upload rate close to the one it announced to the tracker, that leecher will
not be unchoked again for some period of time. In this manner, the possibility of a remote 
leecher initiating a new interaction is left open, yet the benefit from free-riding behavior 
is limited since free-riders will eventually end up choked by most peers.
Since the tracker still returns a random subset of peers, independently of the advertised 
upload capacity, there is no risk of creation of disconnected clusters. In a collaborative 
environment, however, the tracker might even want to return peers based on their capacity, 
as previously proposed~\cite{bharambe06}, in order to speed up cluster creation even more. 
Of course, although the proposed tracker extension is promising, further investigation is 
required to verify that it will work as expected. 

\newpage
\section{Related Work}
\label{related}
There has been a fair amount of work on the performance and behavior of 
BitTorrent systems. 
Bram Cohen, the protocol's creator, has described BitTorrent's main 
mechanisms and their design rationale~\cite{cohen03}. 
There have been several measurement studies examining real BitTorrent traffic.
Izal \textit{et al.}~\cite{izal04} measure several peer characteristics derived
from the tracker log for the Redhat Linux 9 ISO image, including the number of
active peers, the proportion of seeds and leechers, and the geographical spread of peers.
They observe that while there is a correlation between upload and download rates, 
indicating that the choking algorithm is working, the majority of content 
is contributed by only a few leechers and the seeds.
Pouwelse \textit{et al.}~\cite{pouwelse05} study the content availability, integrity,
and download performance for torrents on an once-popular tracker website. 
They observe that the centralized tracker component could potentially be a bottleneck.
Andrade \textit{et al.}~\cite{andrade05} study BitTorrent sharing communities. They
find that sharing-ratio enforcement and the use of RSS feeds to advertise new content
may improve peer contributions, yet torrents with a large number of seeds present
ample opportunity for free-riding.
Furthermore, Guo \textit{et al.}~\cite{guo05} demonstrate that the peer arrival and
departure rate is exponential, and that performance fluctuates widely in small torrents.
Inter-torrent collaboration is proposed as an alternative to providing extra incentives 
for leechers to stay connected after the completion of their download.
A more recent study by Legout \textit{et al.}~\cite{legout06} presents the results
of extensive experiments on real torrents. They show that the rarest-first and choking 
algorithms play a critical role in BitTorrent's performance, and claim that the 
replacement with a volume-based tit-for-tat algorithm, as proposed by other 
researchers~\cite{jun05}, is not appropriate. 
However, they do not identify the reasons behind the properties of the choking 
algorithm and fail to examine its dynamics due to the single-peer viewpoint.

Several analytical studies have formulated models for BitTorrent-like protocols.
Qiu \textit{et al.}~\cite{qiu04} provide a solution to a fluid model of BitTorrent, 
where they study the choking algorithm and its effect on performance. They observe 
that optimistic unchoking may provide a way for peers to free-ride on the system.
Their model assumes peer selection based on global knowledge of all peers in the 
torrent, as well as uniform distribution of pieces. 
Massoulie \textit{et al.}~\cite{massoulie05} introduce a probabilistic model of 
BitTorrent-like systems and argue that overall system performance does not depend 
critically on either altruistic peer behavior or the rarest-first piece selection strategy.  
Fan \textit{et al.}~\cite{fan06} characterize the complete design space of 
BitTorrent-like protocols by providing a model that captures the fundamental trade-off 
between performance and fairness.
Whereas all these models provide valuable insight into the behavior of BitTorrent systems, 
unrealistic assumptions limit their applicability in real scenarios~\cite{guo05, pouwelse05}.

Other researchers have relied on simulations to understand BitTorrent's properties.
Felber \textit{et al.}~\cite{felber04} conducted an initial investigation of the impact 
of different peer arrival rates, peer capacities, and peer and piece selection strategies.
Bharambe \textit{et al.}~\cite{bharambe06} utilize a discrete event simulator to evaluate 
the impact of BitTorrent's core mechanisms and observe that the rate-based tit-for-tat 
strategy is ineffective in preventing unfairness in peer contributions.
They also find that the rarest-first algorithm outperforms alternative piece 
selection strategies. However, they do not evaluate a peer set larger than 15 peers, 
whereas the official implementation has a default value of 80. This may affect the
results since the accuracy of the piece selection strategy is affected by the peer set size. 
Furthermore, Tian \textit{et al.}~\cite{tian06} study peer performance towards the 
end of the download and propose a new peer selection strategy which enables more 
clients to complete their download after the departure of all the seeds.

Researchers have also looked into the feasibility of selfish behavior, when peers attempt
to circumvent BitTorrent mechanisms to gain unfair benefit.
Shneidman \textit{et al.}~\cite{shneidman04} were the first to demonstrate that BitTorrent 
exploits are feasible. They briefly describe an attack to the tracker and an exploit involving 
leechers lying about the pieces they have.
Jun \textit{et al.}~\cite{jun05} argue that the choking algorithm is not sufficient to prevent 
free-riding and propose a new algorithm to enforce fairness in peers' data exchanges.
Liogkas \textit{et al.}~\cite{liogkas07} design and implement three exploits that allow a peer 
who does not contribute to maintain high download rates under specific circumstances. Even though 
such selfish peers can obtain more bandwidth, there is no considerable degradation of the overall
system's quality of service.
Locher \textit{et al.}~\cite{locher06} extend the work in ~\cite{liogkas07} and demonstrate that 
limited free-riding is feasible even in the absence of seeds. They also describe selfish behavior 
in BitTorrent sharing communities.
In addition, Sirivianos \textit{et al.}~\cite{sirivianos07} evaluate an exploit based on maintaining 
a larger-than-normal view of the torrent.
Piatek \textit{et al.}~\cite{piatek07} observe that high-capacity peers typically provide low-capacity 
ones with an unfair share of the data. They design a choking algorithm optimization that reallocates 
the superfluous upload bandwidth to others in order to maximize peer download rates.

Our work differs from all previous studies in its approach and results. We perform the first 
extensive experimental study of BitTorrent in a controlled environment, by monitoring all peers 
in the torrent and examining peer behavior in a variety of scenarios. Our results validate 
protocol properties that have not been previously demonstrated experimentally, and identify new 
properties related to the impact of the initial seed on clustering and sharing incentives.

\section{Conclusion}
\label{conclusion}
In this paper we presented the first experimental investigation of BitTorrent 
systems that links per-peer decisions and overall torrent behavior. Our results 
validate three BitTorrent properties that, though believed to hold, have not 
been previously demonstrated experimentally. We show that the choking algorithm 
enables clustering of similar-bandwidth peers, fosters effective sharing incentives 
by rewarding peers who contribute, and achieves high peer upload utilization for the
majority of the download duration. We also examined the properties of the modified 
choking algorithm in seed state and the impact of initial seed capacity on the 
overall system performance. In particular, we showed that an underprovisioned initial 
seed does not facilitate the clustering of peers and does not provide effective sharing 
incentives. However, even in such a case, the choking algorithm facilitates efficient 
utilization of the available resources by having fast peers help others with their download. 
Based on our observations, we offered guidelines for content providers regarding 
seed provisioning, and discussed a proposed tracker protocol extension that addresses 
an identified limitation of the protocol.

This work opens up many avenues for future research. We are currently developing
an analytical model to express the impact of seed capacity on peer performance.
It would also be interesting to run experiments with the old choking algorithm in 
seed state and compare its properties to the modified algorithm, especially with respect to
the upload of duplicate pieces.
In addition, we would like to investigate the impact of different numbers of 
regular and optimistic unchokes on the protocol's properties. 
It has recently been argued that there is a fundamental trade-off 
between these two kinds of unchokes~\cite{fan06}.
The current values used by the protocol are intuition-based engineering 
choices; we would like to conduct a systematic evaluation of system 
behavior under different parameter values.

\section*{Acknowledgments}
We wish to thank the anonymous reviewers and Michael
Sirivianos for their invaluable feedback.

\bibliographystyle{abbrv}

\end{document}